 \documentclass[keywords,a4paper,11pt]{article}
\usepackage{amsmath}
\usepackage{amssymb}
\usepackage{amsfonts}
\usepackage{latexsym}
\usepackage{epsfig}
\usepackage{float}
\begin{document}

\title{Acoustic response of a rigid frame porous medium slab with a periodic set of inclusions}
\author{J.-P. Groby \thanks{Correspondence to: J.-P.
Groby, CMAP, UMR 7641 CNRS/Ecole Polytechnique, 91128
Palaiseau cedex, France}, A. Wirgin \thanks{LMA, UPR
7051 CNRS, 31 Chemin Joseph-Aiguier, 13402 Marseille cedex 20, France}, and L. de Ryck\thanks{Laboratory of Acoustics and Thermal Physics, Celestijnenlaan 200D, B-3001 Leuven, Belgium}, and W. Lauriks \thanks{Laboratory of Acoustics and Thermal Physics, Celestijnenlaan 200D, B-3001 Leuven, Belgium}}
\maketitle
\begin{abstract}
The acoustic response of a rigid frame porous slab with a periodic set of inclusions is calculated by use of a multipole method. The acoustic properties, in particular the absorption, of such a structure are then derived and studied. Numerical results together with a modal analysis show that the addition of a periodic set of high-contrast inclusions leads to quasi-modes excitation of both the slab and the gratings, and to a large increase of the acoustic absorption of the initial slab, this being partly due to the quasi-modes excitation.
\end{abstract}
\section{Introduction}
This work was initially motivated by the design problem connected
with the determination of the optimal profile of a continuous and/or
discontinuous spatial distribution of the material/geometric
properties of porous materials for the absorption of sound. The equations that
model the acoustic wave propagation in a macroscopically inhomogeneous
rigid frame porous medium were derived in \cite{deryck}. Acoustic
properties of porous materials (foam) suffer from a lack of
absorption particularly at low frequency. The usual way to solve
this problem is by multi-layering \cite{Tanneau}.

In \cite{olny}, the authors considered the reflection of a plane
acoustic wave by a porous slab that presents a periodic set of pits. The medium is homogenized and its behavior
is described in \cite{boutin}. This leads to a drastic increase of
the absorption coeficient at low frequency. In \cite{tournat} the
authors considered the transmission of an acoustic wave through a
porous medium in which randomly-arranged metallic rods are imbedded,
converted, by a procedure called $ISA\beta$, into an equivalent
homogeneous medium which exhibits decreased transmission and
increased absorption.

Periodic arrangements of either surface irregularities or volume
heterogeneities usually lead to energy entrapment either at the
surface or inside the structure, respectively, this being strongly
linked to mode excitation, and to an increase of the absorption
coefficient (first noticed by Wood \cite{wood} and partially
explained by Cutler \cite{cutler}). The particular properties of
such structures have been studied in mechanics, with application
to composite materials \cite{budiansky,hashin,Wu1966}, in optics,
initially motivated by the collection of solar energy
\cite{cuomozieglerwwoodhall,horwitz}, with applications to
photonic crystals \cite{joannopoulos,yablonovitch}, in
electromagnetics, with application to so-called left-handed
materials \cite{ves}, in geophysics, for the study of the
``city-site'' effect \cite{bard,grobyGJIII2006}. The properties of
such structures are now studied to create band-gaps for elastic or
acoustic waves (phononic crystals \cite{Khelif,Laude,Wilm}), but
have only recently been used for the design of sound absorbing or
porous materials \cite{bruijn,kelders1,sapoval1}.

Herein, we  study the influence on the acoustic absorption of the
introduction of a periodic set of fluid-like circular cylinders into
a macroscopically-homogeneous porous slab (the porosity being homogenized
in the equivalent fluid model).
\section{Formulation of the problem}\label{section1}
\begin{figure}[H]
\centering\psfig{figure=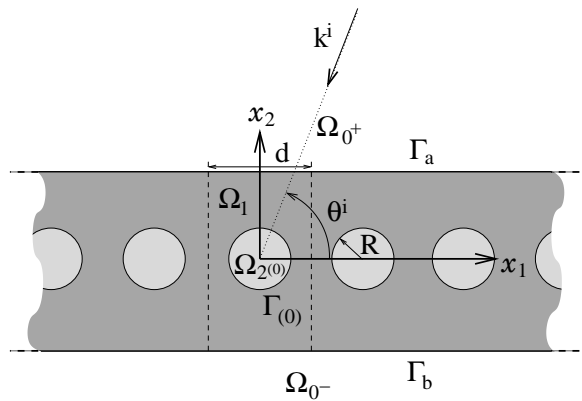,width=8.0cm}
\caption{Sagittal plane representation of the configuration of
plane wave solicitation of a $d$-periodic porous slab with
fluid-like inclusions (of radius $R$) within a porous fluid-like
slab.} \label{section1fig1}
\end{figure}

Both the incident plane acoustic wave and the slab are assumed to be
invariant with respect to the Cartesian coordinate $x_{3}$. A
sagittal $x_{1}\!-\!x_{2}$ plane view of the 2D scattering problem is
given in Fig. \ref{section1fig1}.

Before the introduction of the cylindrical inclusions, the slab is
made of a porous material (e.g., a foam) which is modeled (by
homogenization) as a (macroscopically-homogeneous) equivalent fluid
$M^{1}$. Another equivalent fluid medium $M^{2}$ occupies each
cylindrical inclusion. In the sagittal plane, the $j$th cylinder is
the circular disk $\Omega_{2^{(j)}}$.  The host medium $M^{0}$
occupying the two half spaces $\Omega_{0^{\pm}}$ is air. Thus, we are dealing with a
macroscopically-inhomogeneous slab, the heterogeneity being periodic in
the $x_{1}$ direction with period $d$.

The upper and lower flat, mutually-parallel boundaries of the slab
are $\Gamma_{a}$ and $\Gamma_{b}$. The $x_{2}$
coordinates of these lines are $a$ and $b$, the
thickness $h$ of the slab being $h=a-b$. The circular boundary of
$\Omega_{2^{(j)}}$ is $\Gamma_{(j)}$. The center of the $j=0$ disc
is at the origin $O$ of the laboratory system $Ox_{1}x_{2}x_{3}$.
The union of $\Omega_{0^{+}}$ and $\Omega_{0^{-}}$ is denoted by
$\Omega_{0}$.

The wavevector $\mathbf{k}^{i}$ of the incident plane wave lies in
the sagittal plane and the angle of incidence is $\theta^{i}$
measured counterclockwise from the positive $x_{1}$ axis.
\section{Wave equations}
We designate a total pressure and wavenumber by the generic
symbols $p$ and $k$ respectively, with $p=p^{0^{\pm}},~k=k^{0}$ in
$\Omega_{0^{\pm}}, p=p^{1},~k=k^{1}$ in $\Omega_{1}$ and
$p=p^{2(j)},~k=k^{2}$ in $\Omega_{2^{(j)}}$.

Rather than solving directly for the  pressure $p(\mathbf{x},t)$
(with $\mathbf{x}=(x_{1},x_{2})$), we prefer to deal with
$p(\mathbf{x},\omega)$, related to $p(\mathbf{x},t)$ by the Fourier
transform:
\begin{equation}
p(\mathbf{x},t)=\int_{-\infty}^{\infty}p(\mathbf{x},\omega)e^{-\mbox{i}\omega
t} d\omega~. \label{section1e1bis}
\end{equation}
Henceforth, we drop the $\omega$ in $p(\mathbf{x},\omega)$ so as to
designate the latter by $p(\mathbf{x})$. This function satisfies the
Helmholtz equations
\begin{equation}
\left[ \triangle+(k^{m})^{2}\right]p(\mathbf{x})=0~~;
~~\mathbf{x}\in\Omega_{m}~,~m=0^{\pm},1,2 ~. \label{section1e1ter}
\end{equation}
In an equivalent fluid medium \cite{deryck}, the compressibility and density take the form
\begin{equation}
\begin{array}{ll}
\displaystyle \frac{1}{K_e}=&\displaystyle\frac{\gamma P_0}{\phi\left(\gamma-\left(\gamma-1 \right)\left(1+\mathbf{i}\frac{\omega_c}{\mbox{Pr}^2\omega}G(\mbox{Pr}^2\omega) \right)^{-1} \right)}~,\\
\displaystyle \rho_e=&\displaystyle\frac{\rho_f \alpha_{\infty}}{\phi}\left(1+\mbox{i}\frac{\omega_c}{\omega}F(\omega) \right)~,
\end{array}
\label{section1e1q}
\end{equation}
where $\displaystyle \omega_c=\frac{\sigma \phi}{\rho_f \alpha_{\infty}}$ is the Biot's frequency cut, $\gamma$ the specific heat ratio, $P_0$ the atmospheric pressure, $\mbox{Pr}$ the Prandtl number, $\rho_f$ the density of the fluid in the (interconnected) pores, $\phi$ the porosity, $\alpha_\infty$ the tortuosity, and $\sigma$ the flow resistivity. The correction functions $G(\mbox{Pr}^2\omega)$ \cite{allardchampoux}, $F(\omega)$ \cite{johnson} are given by
\begin{equation}
\begin{array}{ll}
\displaystyle G(\mbox{Pr}^2\omega)=&\displaystyle \sqrt{1-\mbox{i}\frac{\eta \rho_f \alpha_{\infty}^2}{\sigma^2 \phi^2 \Lambda'^2}\mbox{Pr}^2\omega}~,\\
\displaystyle F(\omega)=&\displaystyle \sqrt{1-\mbox{i}\frac{\eta \rho_f \alpha_{\infty}^2}{\sigma^2 \phi^2 \Lambda^2}\omega}.
\end{array}
\label{section1e1c}
\end{equation}
where $\eta$ is the viscosity of the fluid, $\Lambda'$ the thermal characteristic length, and $\Lambda$ the viscous characteristic length.

The incident wave propagates in $\Omega_{0^{+}}$ and is expressed by
\begin{equation}
p^{i}(\mathbf{x})=A^{i}\exp[\mbox{i}(k_{1}^{i}x_{1}-k_{2}^{i}(x_{2}-a)]~,
\label{section1e1fun}
\end{equation}
wherein $k_1^i=-k^0\cos \theta^i$, $k_2^i=k^0\sin \theta^i$ and
$A^{i}=A^{i}(\omega)$ is the signal spectrum.

The new feature, with respect to the canonical case considered in
\cite{grobyMMAS2006}, is the transverse periodicity of
$\displaystyle \cup_{j\in\mathbb{Z}}\Omega_{2^{(j)}}$.

Owing to the plane wave nature of the incident wave, and the
periodic nature of $\displaystyle \cup_{j\in\mathbb{Z}}\Omega_{2^{(j)}}$,  one can show that the field is
quasi-periodic (Floquet theorem), i.e.,
\begin{equation}
p((x_1+nd,x_2))=p((x_1,x_2))e^{\mbox{i}k_1^i nd}\,;\,\forall \mathbf{x}
\in \mathbb{R}^{2}\,;\,\forall n \in \mathbb{Z}~. \label{section1e1}
\end{equation}
Consequently, it suffices to examine the field in the central cell
of the slab which includes the disk $\Omega_{2^{(0)}}$ in order to
obtain the fields, via the Floquet relation, in the other cells.
Henceforth, we adopt the simplified notation:
$\Omega_{2}:=\Omega_{2^{(0)}}$, $\Gamma:=\Gamma_{(0)}$,
$p^{2}=p^{2^{(0)}}$.
\section{Boundary and radiation conditions}\label{section2b}
Since  $M^0$ and $M^1$ are fluid-like, the pressure and the normal
velocity are continuous across the interfaces $\Gamma_{a}$ and
$\Gamma_{b}$:
\begin{equation}
p^{0^+}(\mathbf{x})-p^{1}(\mathbf{x})=0\mbox{, }\forall \mathbf{x} \in \Gamma_a~,
\label{s2be1}
\end{equation}
\begin{equation}
\left(\rho^0\right)^{-1}\partial_n p^{0^+}(\mathbf{x})-
\left(\rho^1\right)^{-1}\partial_n p^{1}(\mathbf{x})=0\mbox{,
}\forall \mathbf{x} \in \Gamma_a~,\label{s2be2}
\end{equation}
\begin{equation}
p^{0^-}(\mathbf{x})-p^{1}(\mathbf{x})=0\mbox{,
}\forall \mathbf{x} \in \Gamma_b~, \label{s2be3}
\end{equation}
\begin{equation}
\left(\rho^0\right)^{-1}\partial_n p^{0^-}(\mathbf{x})-
\left(\rho^1\right)^{-1}\partial_n p^{1}(\mathbf{x})=0\mbox{,
}\forall \mathbf{x} \in \Gamma_b~, \label{s2be4}
\end{equation}
wherein $\mathbf{n}$ denotes the generic unit vector normal to a
boundary and $\partial _n$ designates the operator $\partial
_n=\mathbf{n} \cdot \nabla $.

Since  $M^1$ and $M^2$ are fluid-like,  the pressure and normal
velocity are continuous across the interface $\Gamma$:
\begin{equation}
p^{2}(\mathbf{x})-p^{1}(\mathbf{x})=0\mbox{, }\forall \mathbf{x} \in
\Gamma~, \label{s2be5}
\end{equation}
\begin{equation}
\left(\rho^2\right)^{-1}\partial_n
p^{2}(\mathbf{x})-\left(\rho^1\right)^{-1}\partial_n
p^{1}(\mathbf{x})=0\mbox{, }\forall \mathbf{x} \in \Gamma~.
\label{s2be6}
\end{equation}
The uniqueness of the solution to the  forward-scattering problem is
assured by the radiation conditions :
\begin{equation}
p^{0^+}(\mathbf{x})-p^{i}(\mathbf{x}) \sim \mbox{ outgoing waves ;
}\left|\mathbf{x} \right|\rightarrow \infty~,~x_2>a~, \label{s2be7}
\end{equation}
\begin{equation}
p^{0^-}(\mathbf{x})\sim \mbox{ outgoing waves ; }\forall
\left|\mathbf{x} \right|\rightarrow \infty~,~x_2<b~.
\label{s2be7bis}
\end{equation}
%
\section{Field representations}\label{section2}
Separation of variables, the radiation conditions, and the Floquet
theorem lead to the representations:
\begin{equation}
p^{0^+}(\mathbf{x})=\sum_{p\in\mathbb{Z}}\left[e^{-\mbox{i}k_{2p}^0\left(x_2-a\right)}
\delta_{p0}+R_p
e^{\mbox{i}k_{2p}^0\left(x_2-a\right)}\right]e^{\mbox{i}k_{1p}x_1}
\mbox{, }\forall \mathbf{x}\in \Omega_{0^+}~, \label{s2e1}
\end{equation}
\begin{equation}
p^{0^-}(\mathbf{x})=\sum_{p\in\mathbb{Z}}T_p
e^{\mbox{i}\left(k_{1p}x_1-k_{2p}^0\left(x_2-b\right)\right)}\mbox{,
} \forall \mathbf{x}\in \Omega_{0^-} \label{s2e1b}~,
\end{equation}
wherein $\delta_{p0}$ is the Kronecker symbol, $\displaystyle
k_{1p}=k_1^i+\frac{2p\pi}{d}$,
$k_{2p}^0=\sqrt{(k^0)^2-(k_{1p})^2}$, with
$\Re\left(k_{2p}^0\right)\geq 0$ and $\Im\left(k_{2p}^0\right)\geq
0$.

The field in the central inclusion, with $[\mathbf{r}=(r,\theta)]$, takes the  form
\begin{equation}
p^{2}(\mathbf{r})=\sum_{l\in\mathbb{Z}} C_l J_l\left(k^2
r\right) e^{\mbox{i}l\theta}\mbox{, } \forall (r,\theta) \in
\Omega_2~, \label{s2e5}
\end{equation}
wherein $J_{l}$ is the $l$-th order Bessel function.

It is convenient to combine Cartesian coordinates
$\left(x_1,x_2\right)$ and cylindrical coordinates $(r,\theta)$ to
write the field representation in $\Omega_1$. The latter takes the
form of the sum (by use of the superposition principe) of the
diffracted field by the inclusions $p_d^{1}(\mathbf{x})$ and of
the diffracted field in the slab $p_s^{1}(\mathbf{x})$.
Because of the quasi-periodic aspect of the configuration, the
diffracted field in the slab can be written in Cartesian coordinates as:
\begin{equation}
p_s^{1}(\mathbf{x})=\sum_{p\in\mathbb{Z}}\left(f_{1p}^-
e^{-\mbox{i}k_{2p}^{1}x_2}+f_{1p}^+
e^{\mbox{i}k_{2p}^{1}x_2}\right)e^{\mbox{i}k_{1p} x_1}~.
\label{s2e5b}
\end{equation}
We transform Cartesian to polar coordinates by means of:
\begin{equation}
\begin{array}{ll}
\displaystyle x_1=r\cos\left(\theta\right)~, & \displaystyle
k_{1p}=k^1\cos\left(\theta_p\right)~,\\ 
\displaystyle
x_2=r\sin\left(\theta\right)~, & \displaystyle
k_{2p}^{1}=k^1\sin\left(\theta_p\right)~,
\end{array}
\end{equation}
so that
\begin{equation}
p_s^{1}(\mathbf{r})=\sum_{p\in\mathbb{Z}}\left(f_{1p}^-
e^{\mbox{i}k^1 r \cos(\theta+\theta_p)}+f_{1p}^+ e^{\mbox{i}k^1 r
\cos(\theta-\theta_p)}\right)~.
\end{equation}
Use of the identity
\begin{equation}\label{eidenty}
e^{\mbox{i}k^1 r \cos(\theta-\theta_p)}=\sum_{m\in\mathbb{Z}}
\mbox{i}^m J_m(kr)e^{\mbox{i}m(\theta-\theta_p)}~,
\end{equation}
leads to
\begin{equation}
\begin{array}{ll}
\displaystyle p_s^{1}(\mathbf{r})& \displaystyle =
\sum_{m\in\mathbb{Z}}\sum_{p\in\mathbb{Z}} \mbox{i}^m
\left(f_{1p}^- e^{\mbox{i}m\theta_p}+ f_{1p}^+
e^{-\mbox{i}\theta_p}\right)e^{\mbox{i}m\theta}\\
\displaystyle & \displaystyle =
\sum_{m\in\mathbb{Z}}\sum_{p\in\mathbb{Z}} \left(f_{1p}^-
J_{mp}^-+f_{1p}^+ J_{mp}^+ \right)e^{\mbox{i}m\theta}~,
\end{array}
\end{equation}
wherein $J_{mp}^-=\mbox{i}^m e^{\mbox{i}m\theta_p}$ and
$J_{mp}^+=\mbox{i}^m e^{-\mbox{i}m\theta_p}$.

Let us now introduce: i) $(r^j(P),\theta^j(P))$, the  polar
coordinates of a point P in the system linked to the $j$th cylinder
whose center is at the origin $O_j$, and ii) $(r_l^j,\theta_l^j)$
the polar coordinates of the origin $O_j$ in the polar coordinate
system linked to the $l$th cylinder.

In the general case, the field diffracted  by the inclusion appears
as the sum of the fields diffracted by all the inclusions:
\begin{equation}
p_d^{1}(\mathbf{x})=\sum_{j\in\mathbb{Z}}
\sum_{m\in\mathbb{Z}} B^j H_m^{(1)}\left(k^1 r^j
\right)e^{\mbox{i}m \theta^j}~,
\end{equation}
wherein $H_{m}^{(1)}$ is the first-kind Hankel function of order
$m$.

Using Graf's formula \cite{abramovitz} for the Hankel function leads
to
\begin{multline}
p_d^{1}(\mathbf{r})=\sum_{j\in\mathbb{Z}\cap\{0\}} \sum_{m\in\mathbb{Z}}
\sum_{q\in\mathbb{Z}} J_m^{(1)}\left(k^1 r \right)e^{\mbox{i}m
\theta} H_{m-q}^{(1)}\left(k^1 r_0^j \right)e^{\mbox{i}(q-m)
\theta_0^j}B_q^j \\
+\sum_{m\in\mathbb{Z}} B^0
H_m^{(1)}\left(k^1 r \right)e^{\mbox{i}m \theta}\mbox{, for }R\!\leq\! r\! \leq\! (d-R)~.
\end{multline}
In the case of gratings, $r_0^j=j\times d$, $\theta_0^j=\pi$ if $j<
0$  or $\theta_0^j=0$ if $j\geq 0$, and the quasiperiodicity implies
that the multipole expansion coefficients relative to the $j$th
cylinder of the grating are given by $B_m^j=e^{\mbox{i}
k_i^1}B_m^0=e^{\mbox{i} k_i^1}B_m$, $\forall m \in \mathbb{Z}$.

The field expansion in the vicinity of the central cylinder then
takes the following form (in agreement with \cite{botten}):
\begin{multline}
p_d^{1}(\mathbf{r})=\sum_{l\in\mathbb{Z}} B_l
H_l^{(1)}\left(k^1 r\right)
e^{\mbox{i}l\theta}\\
+\sum_{l\in\mathbb{Z}}J_l\left(k^1r
\right)e^{\mbox{i}l\theta} \sum_{m\in\mathbb{Z}}S_{l-m}B_m \mbox{,
for }R\! \leq\! r\! \leq\! (d-R)~.
 \label{s2e2}
\end{multline}
with $\displaystyle S_{l}=\sum_{j=1}^{\infty} H_{l}^{(1)}\left(k^1 jd \right)
\left[e^{\mbox{i}k_1^i jd}+\left(-1 \right)^l e^{-\mbox{i}k_1^i jd} \right]$.

To derive an alternative form of (\ref{s2e2}) in Cartesian
coordinates, it is more convenient to start from Green's theorem
\begin{equation}
p_d^{1}(\mathbf{r})= \int_\Gamma\left[ \frac{\partial p}{\partial
n}(\mathbf{r^s})G(\mathbf{r}- \mathbf{r^s})- \frac{\partial
G}{\partial n}(\mathbf{r}-\mathbf{r^s})p(\mathbf{r^s})\right]d\mathbf{r^s}~,
\end{equation}
wherein $G$ is the Green's function
 \begin{equation}
G(\mathbf{r}-\mathbf{r^s})=\frac{i}{2d}\sum_{p\in\mathbb{Z}}
\frac{1}{k_{2p}}
e^{\mbox{i}k_{1p}(x_1-x_1^s)+\mbox{i}k_{2p}\left|x_2-x_2^s\right|}~
,
\end{equation}
or, with $[\mathbf{r^s}=(r^s,\theta^s)]$ and for $R \leq x_2$
 \begin{equation}
\begin{array}{ll}
\displaystyle G(\mathbf{r}-\mathbf{r^s})&\displaystyle =
\frac{i}{2d}\sum_{p\in\mathbb{Z}}
\frac{1}{k_{2p}}e^{\mbox{i}k_{1p}x_1+\mbox{i}k_{2p}x_2}
e^{-\mbox{i}k_{1p}x_1^s-\mbox{i}k_{2p}x_2^s}\\
\displaystyle&\displaystyle =\frac{i}{2d}\sum_{p\in\mathbb{Z}}
\frac{1}{k_{2p}} e^{\mbox{i}k_{1p}x_1+\mbox{i}k_{2p}x_2}
e^{-\mbox{i}r^s \cos (\theta^s-\theta_p)}~,\\
\end{array}
\end{equation}
so that with the help of  identity (\ref{eidenty}), we are led to
\begin{equation}
p_d^{1}(\mathbf{x})=\frac{i}{2d}\sum_{p\in\mathbb{Z}}
\frac{1}{k_{2p}}e^{\mbox{i}k_{1p}x_1+\mbox{i}k_{2p}x_2}\times
\mathcal{I}~,
\end{equation}
with
\begin{equation}
\begin{array}{ll}
\displaystyle \mathcal{I}&\displaystyle=-\int_{0}^{2\pi}\!\!kR d\theta
\sum_{m\in\mathbb{Z}} (-\mbox{i})^m J_m\left(kR
\right)e^{\mbox{i}m(\theta_p-\theta)}
\sum_{l\in\mathbb{Z}}\left[B_l
\dot{H}_l^{kR}+\sum_{n\in\mathbb{Z}} S_{l-n}B_n \dot{J}_l(k R)
\right]e^{\mbox{i}l \theta}\\ \displaystyle &\displaystyle
+\int_{0}^{2\pi}\!\!k R d\theta \sum_{m\in\mathbb{Z}} (-\mbox{i})^m
\dot{J}_m\left(kR \right)e^{\mbox{i}m(\theta_p-\theta)}
\sum_{l\in\mathbb{Z}}\left[B_l H_l^{kR}+\sum_{n\in\mathbb{Z}}
S_{l-n}B_n J_l(kR) \right]e^{\mbox{i}l \theta}\\
 &\displaystyle =\frac{-\mbox{i} \pi}{d} \sum_{m\in\mathbb{Z}}(-\mbox{i})^m e^{\mbox{i}m\theta_p} k R \left[J_m(kR)\dot{H}_m^{(1)}(kR)-H_m^{(1)}(kR)\dot{J}_m(kR) \right]B_m\\
&\displaystyle=\frac{2}{d}\sum_{m\in\mathbb{Z}}
(-\mbox{i}^m)e^{\mbox{i}m\theta_p} B_m~,
\end{array}
\end{equation}
wherein we have used the fact that $\displaystyle
J_{-m}(kR)\dot{H}_m^{(1)}(kR)-H_m^{(1)}(kR)\dot{J}_{-m}(kR)=\frac{2\mbox{i}}{\pi
k R}$.

Proceeding in the same way for $-R \geq x_2$ gives:
\begin{equation}
p_d^{1}(\mathbf{x})=\frac{i}{2d}\sum_{p\in\mathbb{Z}}
\frac{1}{k_{2p}} e^{\mbox{i}k_{1p}x_1-\mbox{i}k_{2p}x_2}\times
\mathcal{I}~,
\end{equation}
wherein
\begin{equation}
\begin{array}{ll}
\displaystyle \mathcal{I}&\displaystyle=-\int_{0}^{2\pi}kR d\theta
\sum_{m\in\mathbb{Z}}(-\mbox{i})^m J_m\left(kR
\right)e^{\mbox{i}m(\theta_p+\theta)}
\sum_{l\in\mathbb{Z}}\left[B_l
\dot{H}_l^{kR}+\sum_{m\in\mathbb{Z}} S_{l-m}B_m \dot{J}_l(k R)
\right]e^{\mbox{i}l \theta}\\ \displaystyle &\displaystyle
+\int_{0}^{2\pi}k R d\theta \sum_{m\in\mathbb{Z}} (-\mbox{i})^m
\dot{J}_m\left(kR \right)e^{\mbox{i}m(\theta_p+\theta)}
\sum_{l\in\mathbb{Z}}\left[B_l H_l^{kR}+\sum_{m\in\mathbb{Z}}
S_{l-m}B_m J_l(kR) \right]e^{\mbox{i}l \theta}\\
 &\displaystyle =\frac{-\mbox{i} \pi}{d}\sum_{m\in\mathbb{Z}}(\mbox{i})^m e^{-\mbox{i}m\theta_p} k R \left[J_{-m}(kR)\dot{H}_m^{(1)}(kR)-H_m^{(1)}(kR)\dot{J}_{-m}(kR) \right]B_m\\
&\displaystyle=\frac{2}{d}\sum_{m\in\mathbb{Z}} (-\mbox{i})^m
e^{\mbox{i}m\theta_p} B_m~.
\end{array}
\end{equation}
The field diffracted  by the inclusion is expressed in Cartesian
coordinates by
\begin{equation}
p_d^{1}(\mathbf{x})=\sum_{p\in\mathbb{Z}}
\sum_{l\in\mathbb{Z}} K_{pl}^{\pm} B_l
e^{\mbox{i}\left(k_{1p}x_1\pm k_{2p}^1 x_2\right)}~, \label{s2e3}
\end{equation}
where the signs $+$ and $-$ correspond to $x_2>R$ and $x_2<R$
respectively, and
\begin{equation}
K_{pm}^{+}=\frac{2(-\mbox{i})^m}{dk_{2p}^1}
e^{\mbox{i}m\theta_p}\mbox{, }K_{pm}^{-}=\frac{2(-\mbox{i})^m}{d
k_{2p}^1} e^{-\mbox{i}m\theta_p}~, \label{s2e4}
\end{equation}
with $\theta_p$  such that $k^1 e^{\mbox{i}
\theta_p}=k_{1p}+\mbox{i}k_{2p}^1$, \cite{botten_2,botten}.
\section{Application of the continuity conditions}\label{section3}
Here we  consider only the equations of continuity across the
interfaces $\Gamma_a$ and $\Gamma_b$. The continuity conditions
across $\Gamma$ will be treated in section \ref{section5}.
\subsection{Continuity of the pressure field across $\Gamma_a$}\label{s3ss1}
From (\ref{s2be1}) we obtain
\begin{equation}
\int_{-\frac{d}{2}}^{\frac{d}{2}}
p^{0^+}((x_1,a))e^{-\mbox{i}k_{1l}x_1}dx1-
\int_{-\frac{d}{2}}^{\frac{d}{2}} p^{1}((x_1,a))
e^{-\mbox{i}k_{1l}x_1}dx_1=0\mbox{, } \forall l \in \mathbb{Z}~.
\label{s3ss1e1}
\end{equation}
Introducing the appropriate field representations therein and making use of the orthogonality relation
\begin{equation}
\int_{-\frac{d}{2}}^{\frac{d}{2}}
e^{\mbox{i}\left(k_{1n}-k_{1l}\right)x_1}dx_1= d \delta_{nl}\mbox{,
}\forall (l,n) \in \mathbb{Z}^2~, \label{s3ss1e2}
\end{equation}
gives rise to
\begin{equation}
\delta_{p0}+R_p-f_{1p}^{-}e^{-\mbox{i}k_{2p}^1a}-f_{1p}^{+}e^{\mbox{i}k_{2p}^1a}-\sum_{l\in\mathbb{Z}}
K_{pl}^{+}B_l e^{\mbox{i}k_{2p}^1a}=0~. \label{s3ss1e3}
\end{equation}
%
\subsection{Continuity of the normal component of the velocity across $\Gamma_a$}\label{s3ss2}
From (\ref{s2be2}) we obtain
\begin{multline}
\displaystyle \int_{-\frac{d}{2}}^{\frac{d}{2}}(\rho^0)^{-1} \partial_{x_2}p^{0^+}((x_1,a))
e^{-\mbox{i}k_{1l}x_1}dx1\\
\displaystyle -\int_{-\frac{d}{2}}^{\frac{d}{2}}(\rho^1)^{-1}
\partial_{x_2} p^{1}((x_1,a)) e^{-\mbox{i}k_{1l}x_1}dx_1=0\mbox{, }\forall l \in
\mathbb{Z}~. \label{s3ss2e1}
\end{multline}
Introducing the appropriate field representation therein, and
making use of the orthogonality relation (\ref{s3ss1e2}), gives rise to
\begin{equation}
-\alpha_{p}^{0} \delta_{p0}+\alpha_{p}^{0}
R_p+f_{1p}^{-}\alpha_{p}^{1}
e^{-\mbox{i}k_{2p}^1a}-f_{1p}^{+}\alpha_{p}^{1}e^{\mbox{i}k_{2p}^1a}-
\sum_{l\in\mathbb{Z}} K_{pl}^{+}B_l
\alpha_{p}^{1}e^{\mbox{i}k_{2p}^1a}=0~, \label{s3ss2e2}
\end{equation}
wherein $\displaystyle \alpha_{p}^{i}=\frac{k_{2p}^i}{\rho^i}$, $\forall i=0,1$.
\subsection{Continuity of the pressure field across $\Gamma_b$}\label{s3ss3}
From (\ref{s2be3}) we obtain
\begin{equation}
\int_{-\frac{d}{2}}^{\frac{d}{2}} p^{0^-}((x_1,b))e^{-\mbox{i}k_{1l}x_1}dx1-
\int_{-\frac{d}{2}}^{\frac{d}{2}} p^{1}((x_1,b)) e^{-\mbox{i}k_{1l}x_1}dx_1=
0\mbox{, }\forall l \in \mathbb{Z}~.
\label{s3ss3e1}
\end{equation}
Introducing the appropriate field representation therein, and
making use of the orthogonality relation (\ref{s3ss1e2}), gives rise to
\begin{equation}
T_p-f_{1p}^{-}e^{-\mbox{i}k_{2p}^1b}-f_{1p}^{+}e^{\mbox{i}k_{2p}^1b}-
\sum_{l\in\mathbb{Z}} K_{pl}^{-}B_l e^{\mbox{i}k_{2p}^1b}=0~.
\label{s3ss3e2}
\end{equation}
%
\subsection{Continuity of the normal component of the velocity across $\Gamma_b$}\label{s3ss4}
From (\ref{s2be4}) we obtain
\begin{multline}
\displaystyle \int_{-\frac{d}{2}}^{\frac{d}{2}}(\rho^0)^{-1} \partial_{x_2}p^{0^-}((x_1,b))
e^{-\mbox{i}k_{1l}x_1}dx1\\
\displaystyle -\int_{-\frac{d}{2}}^{\frac{d}{2}}(\rho^1)^{-1}
\partial_{x_2} p^{1}((x_1,b)) e^{-\mbox{i}k_{1l}x_1}dx_1=0\mbox{, }\forall l \in
\mathbb{Z}~. \label{s3ss4e1}
\end{multline}
Introducing the appropriate field representation therein, and making
use of the orthogonality relation (\ref{s3ss1e2}), gives rise to
\begin{equation}
-\alpha_{p}^{0}T_p+f_{1p}^{-}\alpha_{p}^{1}e^{-\mbox{i}k_{2p}^1b}-f_{1p}^{+}
\alpha_{p}^{1}e^{\mbox{i}k_{2p}^1b}+\sum_{l\in\mathbb{Z}}
K_{pl}^{-}B_l\alpha_{p}^{1}e^{\mbox{i}k_{2p}^1b}=0~.
\label{s3ss4e2}
\end{equation}
%
\section{Determination of the unknowns}\label{section5}
From (\ref{s3ss1e2}), (\ref{s3ss2e2}), (\ref{s3ss3e2}) and (\ref{s3ss4e2}) we get the expressions of $f_{1p}^{-}$ and $f_{1p}^{+}$ in terms of $B_l$. Introducing the latter into (\ref{s2e5b}) together with (\ref{s2e2}) leads to
\begin{multline}
p^{1}(\mathbf{x})=\sum_{l\in\mathbb{Z}} B_l
H_l^{(1)}\left(k^1
r\right)e^{\mbox{i}l\theta}+\sum_{l\in\mathbb{Z}}J_l\left(k^1r
\right)e^{\mbox{i}l\theta}\sum_{m\in\mathbb{Z}}S_{l-m}B_m \\
+\sum_{p\in\mathbb{Z}}\left(F_{1p}^-
e^{-\mbox{i}k_{2p}^{1}x_2}+F_{1p}^+
e^{\mbox{i}k_{2p}^{1}x_2}\right)e^{\mbox{i}k_{1p }x_1}\\
-\sum_{p\in\mathbb{Z}} \sum_{l\in\mathbb{Z}} B_l
\frac{e^{\mbox{i}k_{2p}^1L}\left(\alpha_{p}^0-\alpha_{p}^1
\right)^2}{D_p}\left(K_{pl}^{+}e^{\mbox{i}\left(k_{1p}x_1+k_{2p}^1
x_2 \right)}+K_{pl}^{-}e^{\mbox{i}\left(k_{1p}x_1-k_{2p}^1 x_2
\right)} \right)\\ +\sum_{p\in\mathbb{Z}} \sum_{l\in\mathbb{Z}}
B_l
\frac{\left(\left(\alpha_{p}^0\right)^2-\left(\alpha_{p}^1\right)^2
\right)}
{D_p}\\
\times\left(K_{pl}^{+}e^{\mbox{i}k_{2p}^1\left(a+b\right)}e^{\mbox{i}\left(k_{1p}x_1-k_{2p}^1
x_2 \right)}+
K_{pl}^{-}e^{-\mbox{i}k_{2p}^1\left(a+b\right)}e^{\mbox{i}\left(k_{1p}x_1+\mbox{i}k_{2p}^1
x_2 \right)} \right)~, \label{s5e1}
\end{multline}
wherein
\begin{equation}
\begin{array}{l}
\displaystyle D_p=2\mbox{i}\sin\left(k_{2p}^1 L
\right)\left(\left(\alpha_{p}^0\right)^2+\left(\alpha_{p}^1\right)^2
\right)-4 \alpha_{p}^0\alpha_{p}^1 \cos\left(k_{2p}^1 L \right)~,\\
\displaystyle
F_{1p}^-=-\frac{2\alpha_{p}^0\left(\alpha_{p}^0+\alpha_{p}^1
\right) }{ D_p}e^{\mbox{i}k_{2p}^1b}\delta_{p0}~,\\ \displaystyle
F_{1p}^+=\frac{2\alpha_{p}^0\left(\alpha_{p}^0-\alpha_{p}^1
\right) }{ D_p} e^{-\mbox{i}k_{2p}^1b}\delta_{p0}~.
\end{array}
\end{equation}
To proceed further, we need to convert the cartesian form to the
cylindrical harmonic form:
\begin{multline}
p^{1}(\mathbf{r})=\sum_{l\in\mathbb{Z}} B_l
H_l^{(1)}\left(k^1
r\right)e^{\mbox{i}l\theta}+\sum_{l\in\mathbb{Z}} J_l\left(k^1r
\right)e^{\mbox{i}l\theta}\sum_{m\in\mathbb{Z}} S_{l-m}B_m \\ +
\sum_{l\in\mathbb{Z}} \sum_{p\in\mathbb{Z}}
\left(J_{lp}^{-}F_{1p}^- +J_{lp}^{+} F_{1p}^ +\right) J_l(k^1r)
e^{\mbox{i}l \theta}\\
+\sum_{l\in\mathbb{Z}} J_l\left(k^1r
\right)e^{\mbox{i}l\theta} \sum_{n\in\mathbb{Z}}
\sum_{p\in\mathbb{Z}} \left(Q_{lnp}-P_{lnp} \right)B_n~,
\label{s5e2}
\end{multline}
wherein
\begin{equation}
\begin{array}{l}
\displaystyle Q_{lnp}=\frac{\left(\alpha_{p}^0\right)^2-\left(\alpha_{p}^1\right)^2}{k_{2p}^1 D_p} \times \frac{4(-\mbox{i})^{l-n}}{d} \cos\left(k_{2p}^1\left(a+b \right)+\left(l+n \right)\theta_p \right)~,\\
\displaystyle P_{lnp}=\frac{e^{\mbox{i}k_{2p}^1 L}
\left(\alpha_{p}^0- \alpha_{p}^1 \right)^2}{k_{2p}^1 D_p}\times
\frac{4(-\mbox{i})^{l-n}}{d} \cos\left((l-n)\theta_p \right)~.
\end{array}
\end{equation}
Central to the multipole method are the local field expansions or
multipole expansions  around each inclusion
\cite{botten_2,maystre,botten}. Because $p^{1}(\mathbf{r})$
satisfies a Helmholtz equation inside and outside the cylinder of
the unit cell, in the vicinity of the cylinder we can write
\begin{equation}
p^{1}(\mathbf{r})=\sum_{l\in\mathbb{Z}} B_l
H_l^{(1)}\left(k^1 r\right)
e^{\mbox{i}l\theta}+\sum_{l\in\mathbb{Z}} J_l\left(k^1r
\right)e^{\mbox{i}l\theta} A_l~. \label{s5e3}
\end{equation}
By identifying (\ref{s5e2}) with (\ref{s5e3}), we find
\begin{equation}
A_l=\sum_{m\in\mathbb{Z}} S_{l-m}B_m+\sum_{p\in\mathbb{Z}}
\left(J_{lp}^{-}F_{1p}^- + J_{lp}^{+} F_{1p}^
+\right)+\sum_{m\in\mathbb{Z}} \sum_{p\in\mathbb{Z}}
\left(Q_{lmp}-P_{lmp} \right)B_m~. \label{s5e4}
\end{equation}
At this point, we account for the two equations (\ref{s2be6})  and
(\ref{s2be7}). It is well-known that the coefficients of the
scattered field and those of the locally incident field are linked
by a matrix relation depending on the parameters of the cylinder
only, i.e.,
\begin{equation}
B_l=\frac{\gamma^1 \dot{J}_l(k^1 R)J_l(k^2 R)-\gamma^2 \dot{J}_l(k^2 R)J_l(k^1 R)}{\gamma^2 \dot{J}_l(k^2 R)H_l^{(1)}(k^1 R)-\gamma^1 \dot{H}_l^{(1)}(k^1 R)J_l(k^2 R)} A_l=R_l A_l~,
\label{s5e5}
\end{equation}
wherein $\displaystyle \gamma^j=\frac{k^j}{\rho^j}$. Denoting $\mathbf{B}$ the infinite column matrix of components $B_l$, (\ref{s5e4}) together with (\ref{s5e5}) may be written in the matrix form
\begin{equation}
\left(\mathbf{I}-\mathbf{R}\mathbf{S}-\mathbf{R}\left(\mathbf{Q}-
\mathbf{P} \right) \right)\mathbf{B}=\mathbf{R}\mathbf{F}~.
\label{s5e6}
\end{equation}
with $\mathbf{F}$ the column matrix of $m$th element $\displaystyle \sum_{p\in \mathbb{Z}} J_{mp}^- F_{1p}^-+J_{mp}^+ F_{1p}^+$, $\mathbf{I}$ the identity matrix, $\mathbf{R}$ the diagonal matrix of component $R_l$ and $\mathbf{S}$, $\mathbf{Q}$ and $\mathbf{P}$ three square matrices of respective $(m,q)$th element $S_{m-q}$, $\displaystyle \sum_{p\in \mathbb{Z}} Q_{mqp}$ and $\displaystyle \sum_{p\in \mathbb{Z}} P_{mqp}$.\\
\newline
{\textbf{Remark:}} In case of a Neumann type boundary condition, the relation (\ref{s5e5}) takes the form
\begin{equation}
B_l=\frac{- \dot{J}_l(k^1 R)}{ \dot{H}_l^{(1)}(k^1 R)}
A_l:=R_l^{\mathcal{N}} A_l~. \label{s5e5bis}
\end{equation}
%
\section{Evaluation of the transmitted and reflected fields}\label{section6}
Once  (\ref{s5e6}) is solved for $B_l$, $\forall l \in \mathbb{Z}$,
we can derive, from (\ref{s3ss1e3}) and (\ref{s3ss3e2}), expressions
for $R_p$ and $T_p$ depending on $B_l$, $\forall l \in \mathbb{Z}$,
which, after introduction into (\ref{s2e1}), leads to the expression
of the pressure field in $\Omega_{0^+}$:
\begin{multline}
p^{0^+}(\mathbf{x})=e^{\mbox{i}\left(k_{1}^ix_1-k_{2}^{0i}\left(x_2-a
\right)\right)}+\frac{\mbox{i}\sin\left(k_{2}^{1i}L\right)\left((\alpha^{0i})^2-(\alpha^{1i})^2\right)e^{\mbox{i}\left(k_{1}^ix_1+k_{2}^{0i}\left(x_2-a
\right)\right)}}{\mbox{i}\left((\alpha^{0i})^2+(\alpha^{1i})^2\right)
\sin\left(k_{2}^{1i}L
\right)-2\alpha^{1i}\alpha^{0i}\cos\left(k_{2}^{1i}L
\right)}\\ +\sum_{p\in\mathbb{Z}}
\sum_{l\in\mathbb{Z}}\frac{4(-\mbox{i}^l)}{d k_{2p}^1}
\frac{1}{\mbox{i}\left((\alpha_p^{0})^2+
(\alpha_p^{1})^2\right) \sin\left(k_{2p}^{1}L
\right)-2\alpha_p^{1}\alpha_p^{0}\cos\left(k_{2p}^{1}L \right)}\\
\times\alpha_p^1\left( -\mbox{i}\alpha_p^0
\sin\left(l\theta_p+k_{2p}^1 b \right) - \alpha_p^1 \cos\left(
l\theta_p+k_{2p}^1 b \right)\right)e^{\mbox{i}\left(k_{1p} x_1
+k_{2p}^0\left(x_2-a
\right)\right)}~,
\end{multline}
and, after introduction into (\ref{s2e1b}), leads to

\begin{multline}
p^{0^-}(\mathbf{x})=\frac{-2\alpha^{1i} \alpha^{0i}
e^{\mbox{i}\left(k_{1}^ix_1-k_{2}^{0i}\left(x_2-b
\right)\right)}}{\mbox{i}\left((\alpha^{0i})^2+(\alpha^{1i})^2\right)
\sin\left(k_{2}^{1i}L
\right)-2\alpha^{1i}\alpha^{0i}\cos\left(k_{2}^{1i}L
\right)}\\ +\sum_{p\in\mathbb{Z}} \sum_{l\in\mathbb{Z}}
\frac{4(-\mbox{i}^l)}{d k_{2p}^1} \frac{1}
{\mbox{i}\left((\alpha_p^{0})^2+(\alpha_p^{1})^2\right)
\sin\left(k_{2p}^{1}L
\right)-2\alpha_p^{1}\alpha_p^{0}\cos\left(k_{2p}^{1}L \right)}\\
\times \alpha_p^1
\left(\mbox{i}\alpha_p^0 \sin\left(l\theta_p+k_{2p}^1 a \right) -
\alpha_p^1 \cos\left( l\theta_p+k_{2p}^1 a \right)\right)
e^{\mbox{i}\left(k_{1p} x_1 -k_{2p}^0\left(x_2-b \right)\right)}
~.
\end{multline}
%
\section{Evaluation of pressure field in the slab}\label{section7}
We  obtain from (\ref{s5e1}) the final expression of the field in
$\Omega_1$:
\begin{multline}
p^{1}(\mathbf{x})=\frac{2\alpha^{0i}\left(\mbox{i}\alpha^{0i}
\sin\left(k_{2}^{1i}(x_2-b) \right)-\alpha^{1i}
\cos\left(k_{2}^{1i}(x_2-b) \right)\right)}{
\mbox{i}\left((\alpha^{0i})^2+(\alpha^{1i})^2\right)
\sin\left(k_{2}^{1i}L
\right)-2\alpha^{1i}\alpha^{0i}\cos\left(k_{2}^{1i}L \right)   }\\
+\sum_{l\in\mathbb{Z}} B_l H_l^{(1)}\left(k^1
r\right)e^{\mbox{i}l\theta}+\sum_{l\in\mathbb{Z}}J_l\left(k^1r
\right)e^{\mbox{i}l\theta}\sum_{m\in\mathbb{Z}}S_{l-m}B_m\\
-\sum_{p\in\mathbb{Z}} \sum_{l\in\mathbb{Z}}
B_l\frac{2(-\mbox{i}^l)}{dk_{2p}^1} \frac{1}{\mbox{i}
\left((\alpha_p^{0})^2+(\alpha_p^{1})^2\right)
\sin\left(k_{2p}^{1}L \right)-2\alpha_p^{1}
\alpha_p^{0}\cos\left(k_{2p}^{1}L \right)}\\
\times \cos(k_{2p}^1
x_2+l\theta_p)
e^{\mbox{i}k_{2p}^1L}\left(\alpha_{p}^0-\alpha_{p}^1
\right)^2+\left((\alpha_p^1)^2- (\alpha_p^0)^2
\right)\cos(k_{2p}^1\left(x_2-(a+b) \right) -l \theta_p)~. \label{s7e1}
\end{multline}
Use of the continuity conditions on $\Gamma$ leads to:
\begin{equation}
p^{2}(\mathbf{r})=\sum_{l\in\mathbb{Z}}\frac{\gamma^2
\dot{J}_l\left(k^2 R \right) J_l\left(k^1 R \right)-\gamma^1
\dot{J}_l\left(k^1 R \right) J_l\left(k^2 R \right)}
{\gamma^1\left( \dot{H}_{l}^{(1)}\left(k^1 R \right) J_l\left(k^1
R \right)- \dot{J}_l\left(k^1 R \right) H_{l}^{(1)}\left(k^1 R
\right) \right)}B_l J_l\left(k^2 r \right) e^{\mbox{i}l\theta}~.
\label{s7e2}
\end{equation}

{\textbf{Remark :}} The fields in $\Omega_j$, $j=0^+\mbox{,
}1\mbox{, }0^-$ are the sum of i) the field in absence of the
inclusions (whose expressions are the same as those  in
\cite{grobyMMAS2006}) with ii) the field due to the presence of the
inclusions.

{\textbf{Remark :}} The field due to the presence of the inclusions,
when compared with the Green's function as calculated in
\cite{grobyMMAS2006} in the case of a line source located in the
slab, takes the form of the field radiated by induced periodic
sources. The latter do not add energy to the system, but rather
entail a redistribution of the energy in the frequency range of the
solicitation.
\section{Modal analysis}\label{section8b}
%
\subsection{Modal analysis without inclusions}\label{section8bss1}
In the absence of inclusions, the resolution of the problem reduces
to:
\begin{equation}
\left[
\begin{array}{ll}
\displaystyle \left(\alpha^{1i}-\alpha^{0i} \right)& \displaystyle -e^{-\mbox{i}k_2^{0i}L}\left(\alpha^{1i}+\alpha^{0i} \right)\\[8pt]
\displaystyle \left(\alpha^{1i}+\alpha^{0i} \right)&\displaystyle -e^{\mbox{i}k_2^{0i}L}\left(\alpha^{1i}-\alpha^{0i} \right)
\end{array}
\right]
\left[
\begin{array}{l}
\displaystyle T\\[8pt]
\displaystyle R
\end{array}
\right]=
\left[
\begin{array}{l}
\displaystyle e^{-\mbox{i}k_2^{0i}L}\left(\alpha^{1i}-\alpha^{0i} \right)\\[8pt]
\displaystyle e^{\mbox{i}k_2^{0i}L}\left(\alpha^{1i}+\alpha^{0i} \right)
\end{array}
\right]~. \label{section8bss1e1}
\end{equation}
The natural frequencies modes  of the configuration are obtained by
turning off the excitation. The resulting matrix equation possesses
a non-trivial solution only if the determinant of the matrix
vanishes, i.e. :
\begin{equation}
\mbox{i}\left((\alpha^{0i})^2+(\alpha^{1i})^2\right)
\sin\left(k_{2}^{1i}L \right)-
2\alpha^{1i}\alpha^{0i}\cos\left(k_{2}^{1i}L \right)=0~.
\label{section8bss1e2}
\end{equation}
\begin{figure}[H]
\begin{minipage}{7.0cm}
\centering\psfig{figure=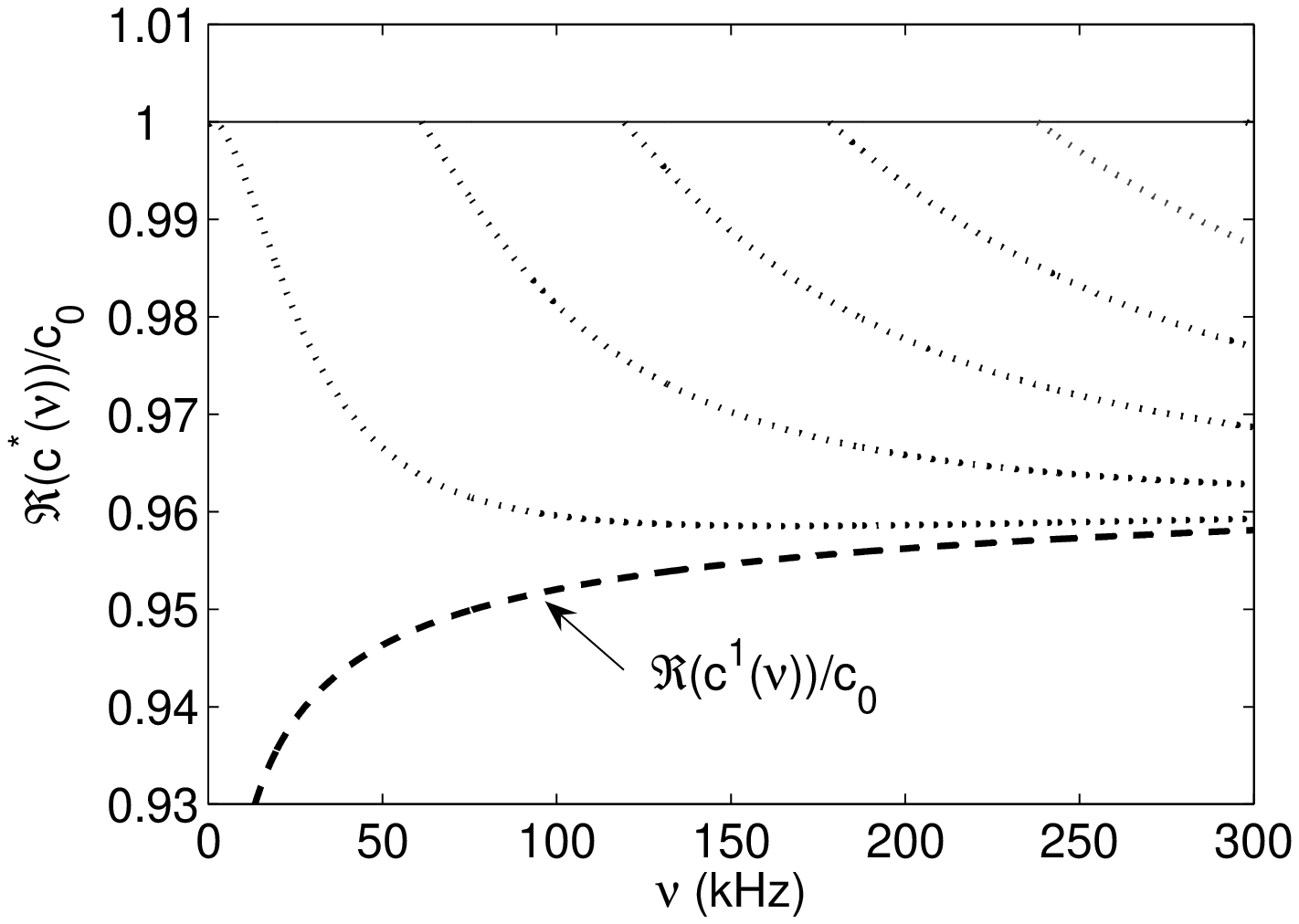,width=7.0cm}
\end{minipage}\hfill
\begin{minipage}{7.0cm}
\centering\psfig{figure=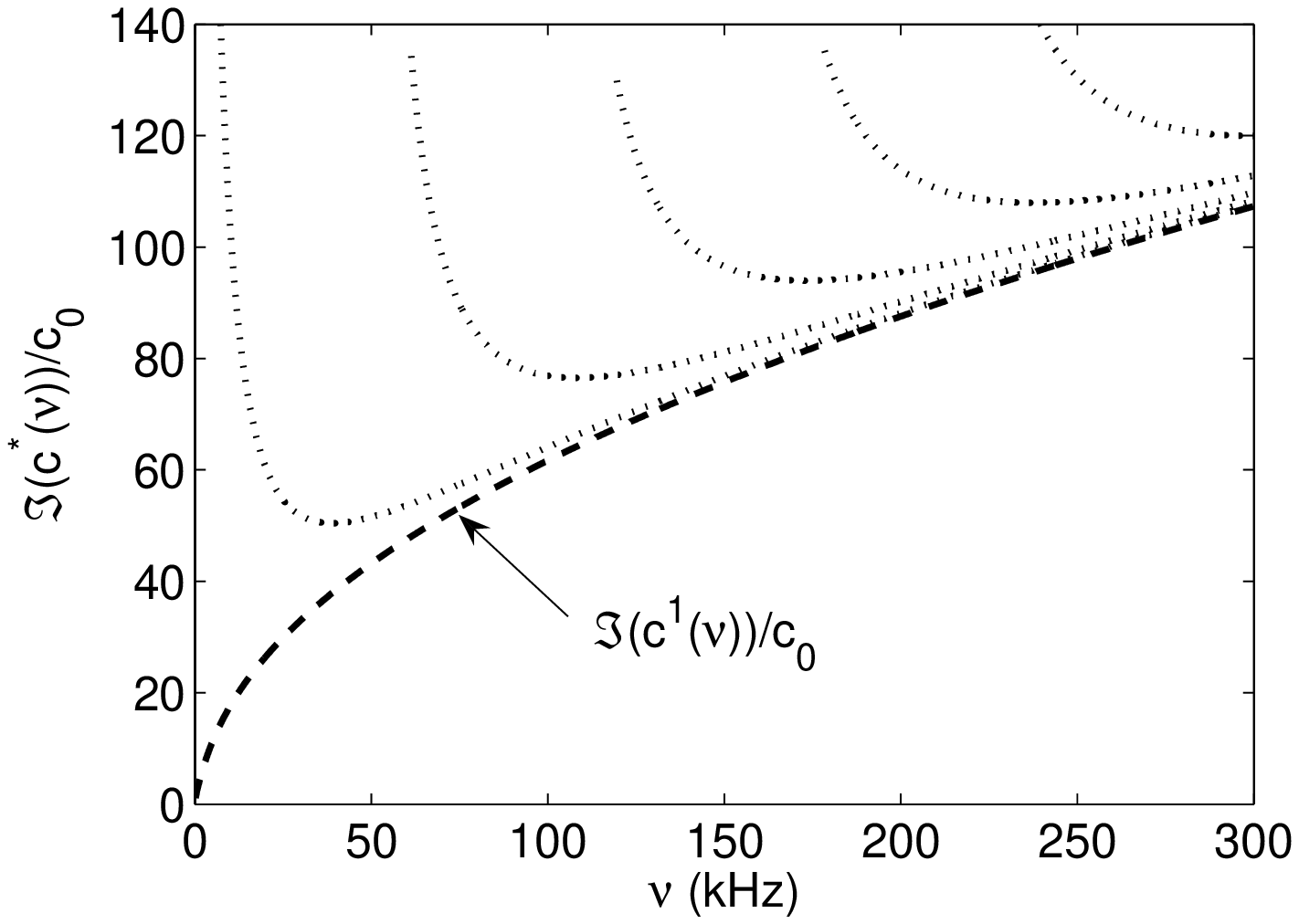,width=7.0cm}
\end{minipage}
\caption{Real part (left panel) and imaginary part (right panel) of the roots of the dispersion relation.}
\label{section8bss1f1}
\end{figure}

Fig.  \ref{section8bss1f1} depicts the real and the imaginary parts
of the solution $c^{\star}(\omega)=\omega/(k_1^{\star})$ of the
previous equation with the mechanical characteristics used in the
section \ref{section9}. To solve the latter, we proceed as in
\cite{wirginandgrobygji2005I}. The dispersion relation cannot vanish
when the incident  wave takes the form of a plane wave, because we
must have $\Re\left(k_1^{\star}\right)\geq k^0$, whereas $k_1^i\in
[-k_0,k_0]$ in case of a incident plane wave.
\subsection{Modal analysis with inclusions}\label{section8bss2}
For the slab with periodic inclusions,  the problem reduces to the
resolution of the linear system (\ref{s5e6}). As
previously, the natural frequencies of the modes of the
configuration are obtained by turning off the excitation, embodied
in the vector $\mathbf{R} \mathbf{F}$. The resulting equation
possesses a non-trivial solution only if the determinant of the
matrix vanishes:
\begin{equation}
\mbox{det}\left(\mathbf{I}-\mathbf{R}
\mathbf{S}-\mathbf{R}\left(\mathbf{Q}-\mathbf{P} \right) \right)=0~.
\label{section8bss2e1}
\end{equation}
A procedure, called the partition method, for solving this equation,
is not easy to apply because the off-diagonal elements of the matrix
are not small compared to the diagonal elements. Even at low
frequency (i.e. $\Re(k^1) R<\!\!<1$), $B_j=B_j^{(2)} \times \left(k^1
R \right)^2 + O \left(\left(k^1 R\right)^2 \right)$ for $j=\{-1, 0
,1 \}$, at least three terms should be taken into account.

An iterative scheme can be employed to solve (\ref{s5e6}) and obtain an
approximate dispersion relation. We re-write this equation in the
form
\begin{multline}
\left(1-R_l \left (S_{0}+ \sum_{p\in\mathbb{Z}} \left(
Q_{llp}-P_{llp} \right) \right) \right)B_l=\\ R_l
\sum_{m\in\mathbb{Z}} \left( S_{l-m}+ \sum_{p\in\mathbb{Z}}
\left(Q_{lmp}-P_{lmp} \right)\right) B_m \left(1-\delta_{ml}
\right)\\
+R_l \sum_{p\in\mathbb{Z}} \left(J_{lp}^{-}F_{1p}^-
+J_{lp}^{+} F_{1p}^+\right)\mbox{, }\forall l \in \mathbb{Z}~.
\label{section8bss2e2}
\end{multline}
The iterative procedure for solving this linear set of equations is:
\begin{equation}
\left\{
\begin{array}{l}
\displaystyle B_l^{(0)}= R_l \frac{\displaystyle
\sum_{p\in\mathbb{Z}} \left(J_{lp}^{-}F_{1p}^- +J_{lp}^{+}
F_{1p}^+\right)}{\displaystyle 1-R_l \left (S_{0}+
\sum_{p\in\mathbb{Z}} \left( Q_{llp}-P_{llp} \right)
\right)}\\ \displaystyle B_l^{(n)}= R_l \frac{1}{\displaystyle 1-R_l \left (S_{0}+ \sum_{p\in\mathbb{Z}} \left(
Q_{llp}-P_{llp} \right) \right)}\times
\left[\sum_{p\in\mathbb{Z}}
\left(J_{lp}^{-} F_{1p}^- +J_{lp}^{+} F_{1p}^+\right)
\right.\\
\displaystyle \left. +\sum_{m\in\mathbb{Z}} \left( S_{l-m}+ \sum_{p\in\mathbb{Z}}
\left(Q_{lmp}-P_{lmp} \right)\right) B_m^{(n-1)}
\left(1-\delta_{ml} \right)
\right]
\mbox{, } \forall n\in
\mathbb{N}^{*}~,
\end{array}
\right.
\end{equation}
from which it becomes apparent that the solution of $B_l^{(n)}$, to
any order of approximation, is expressed as a fraction, the
denominator of which (not depending on the order of approximation),
can become large for certain couples $(k_{1p},\omega)$ so as to make
$B_l^{(n)}$, and (possibly) the field large for these values.

When this happens, a natural mode of the configuration, comprising
the inclusions and the slab, is excited, this taking the form of a
resonance with respect to $B_l^{(n)}$, i.e. with respect to a plane
wave component of the field in the slab relative to the inclusions.
As $B_l^{(n)}$ is related to $f_{1p}^+$, $f_{1p}^-$, $T_p$ and
$R_{p}$, the structural resonance manifests itself for the same
$(k_{1p},\omega)$ as concerns the field in the slab and in the air.
\begin{multline}
1-R_l \left (S_{0}+ \sum_{p\in\mathbb{Z}} \left( Q_{llp}-P_{llp}
\right) \right)= 1-R_l\left( \sum_{j=1}^{\infty}
H_0^{(1)}\left(k^1 j d \right) 2\cos \left(k_{1}^i jd \right)
\right. \\  +\sum_{p\in\mathbb{Z}}\frac{2}{d k_{2p}^1}\frac{1}{
\left(\mbox{i}\sin\left(k_{2p}^1 L
\right)\left(\left(\alpha_{p}^0\right)^2+\left(\alpha_{p}^1\right)^2
\right)-2 \alpha_{p}^0\alpha_{p}^1 \cos\left(k_{2p}^1 L \right) \right)}\\
\left.\times \left(\cos\left(k_{2p}^1 \left(a+b \right)+2 l \theta_p
\right)(-1)^l \left((\alpha_p^0)^2- (\alpha_p^1)^2
\right)-e^{\mbox{i}k_{2p}^1L}\left(\alpha_p^0-\alpha_p^1 \right)^2\right)\right)=0
~.
\end{multline}
The latter equation is the sum of a term linked to the  grating
embodied in $1-R_l S_{0}$ with a term linked to the slab embodied
in $-R_l\sum_{p\in\mathbb{Z}} \left( Q_{llp}-P_{llp} \right)$. This
can be interpreted as a perturbation of the dispersion relation of
the gratings by the presence of the slab.

\section{Numerical results}\label{section9}
The ambiant and saturating fluid is the air medium ($\rho_0=\rho_f=1.213 kg.m^{-3}$,
$\displaystyle c_0=\sqrt{\frac{\gamma P_0}{\rho_0}}$, with $P_0=1.01325 \times
10^5 Pa$ and $\gamma=1.4$, $\eta=1.839 \times 10^{-5} kg.m^{-1}.s^{-1}$).  The infinite layer is $1 \times 10^{-2}m$ thick and filled
with a polymer foam $M^1$. The radius of the inclusions is constant
equal to $R=2.5 \times 10^{-3}m$. We vary the center-to-center
distance between each inclusion from $d=1\times 10^{-2}m$ to
$d=2.5\times 10^{-2}m$. The inclusions are either filled with the air
medium (which define the so-called type 2 sample), with a
melamin-foam (which define the so-called type 1 sample) or with such
a material that the condition upon $\Gamma$ is the Neumann one
(which define the so-called type 3 sample).

The medium $M^1$ is characterized by $\phi=0.96$, $\alpha_\infty=1.07$,
$\Lambda=273 \times 10^{-6}m$, $\Lambda'=672 \times 10^{-6}m$,
$\sigma=2843 N.s.m^{-4}$, while the melamine-foam is characterized by
$\phi=0.99$, $\alpha_\infty=1.001$, $\Lambda=150 \times 10^{-6}m$,
$\Lambda'=150 \times 10^{-6}m$, $\sigma=12 \times 10^3 N.s.m^{-4}$. The incident angle is $\theta^i=0$.

 The infinite sum $\displaystyle \sum_{m\in\mathbb{Z}}$ over the indices of the  modal representation of the
  diffracted field by a cylinder is truncated as $\displaystyle \sum_{m=-M}^{M}$ such that
\begin{equation}
\displaystyle M=\mbox{int}\left(\Re\left(4.05 \times \left(k^1 R \right)^{\frac{1}{3}}+k^1 R \right) \right)+10~.
\end{equation}
On the other hand, the infinite sum $\displaystyle
\sum_{p\in\mathbb{Z}}$ over the indices of the $k_{1p}$ is found
to depend on the frequency and on the period of the grating. We
also use an empirical rule we have determined by performing a large number of numerical experiments $\displaystyle \sum_{p=-P}^{P}$ such
that
\begin{equation}
\displaystyle P=\frac{d \Re\left(k^1 \right)}{2 \pi}+50\mbox{int}\left(e^{\displaystyle \frac{-\omega}{2 \pi 50 \times 10^{3} }}\right)+10\left(- \mbox{ln}\left(\frac{d}{25 \times 10^{-3}} \right)\right)+5~.
\end{equation}
In the latter equations $\mbox{int}\left(a\right)$ represents the integer part of $a$.

\begin{figure}[H]
\begin{minipage}{4.5cm}
\centering\psfig{figure=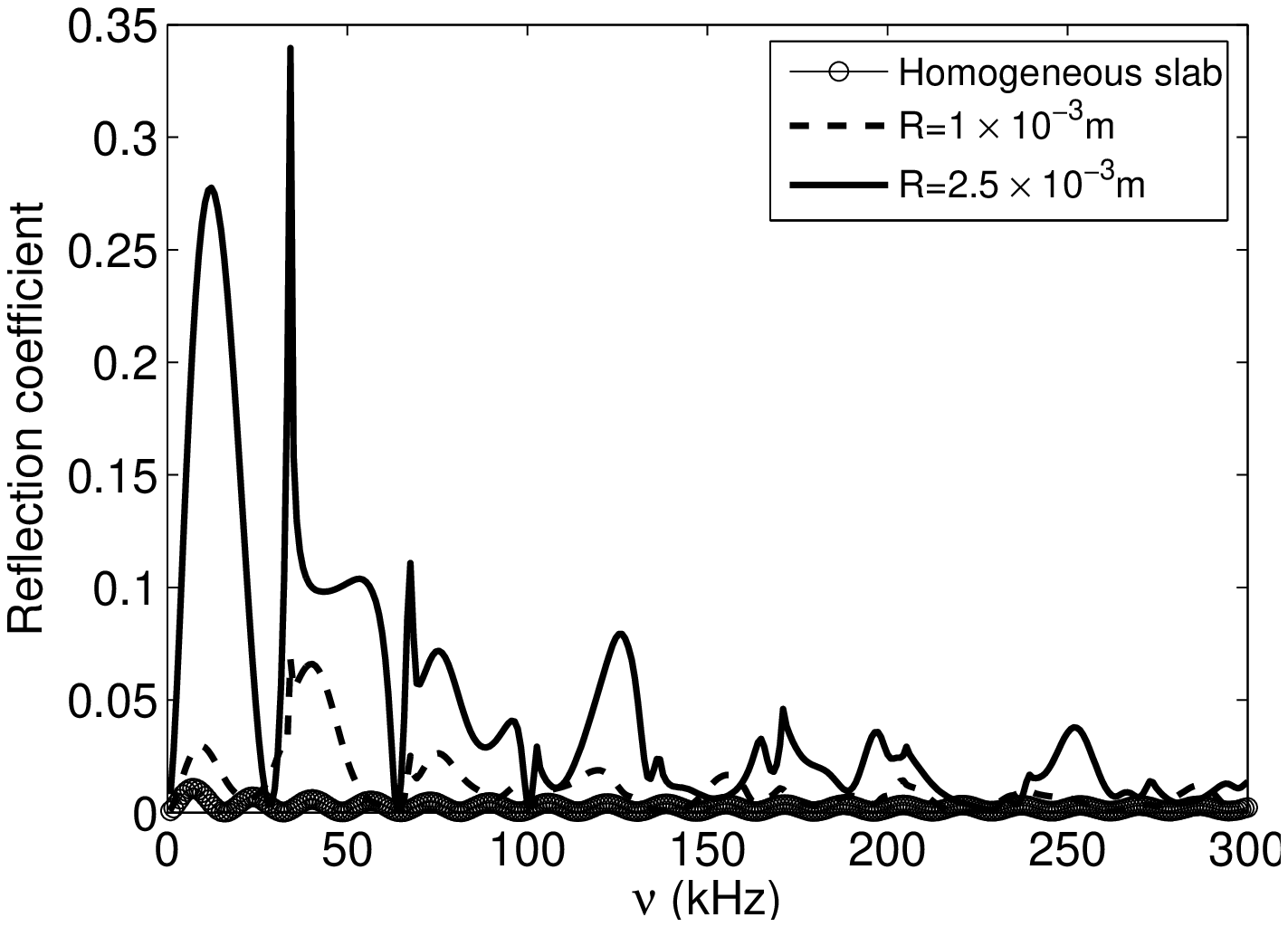,width=4.7cm}
\end{minipage}\hfill
\begin{minipage}{4.5cm}
\centering\psfig{figure=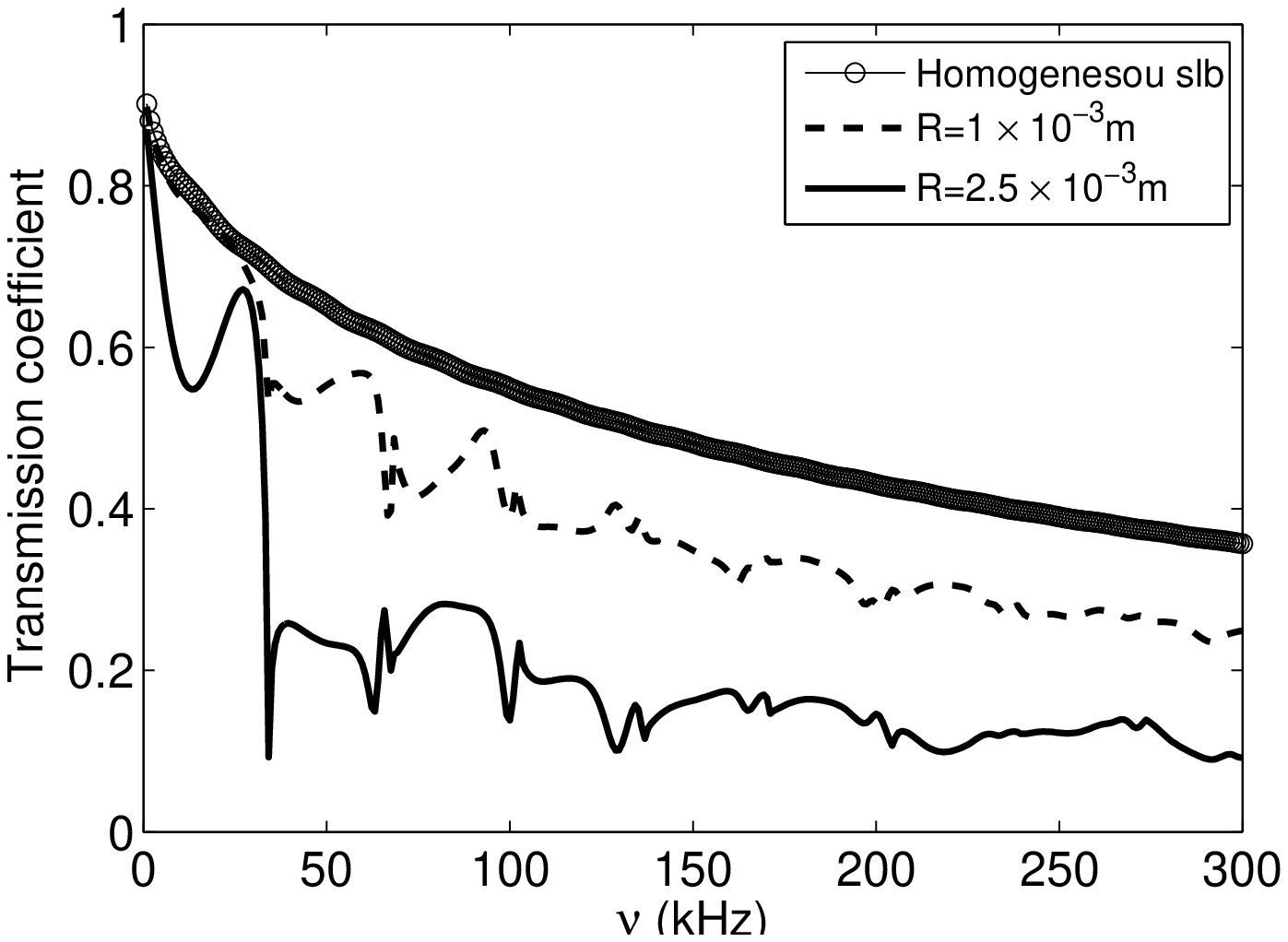,width=4.7cm}
\end{minipage}\hfill
\begin{minipage}{4.5cm}
\centering\psfig{figure=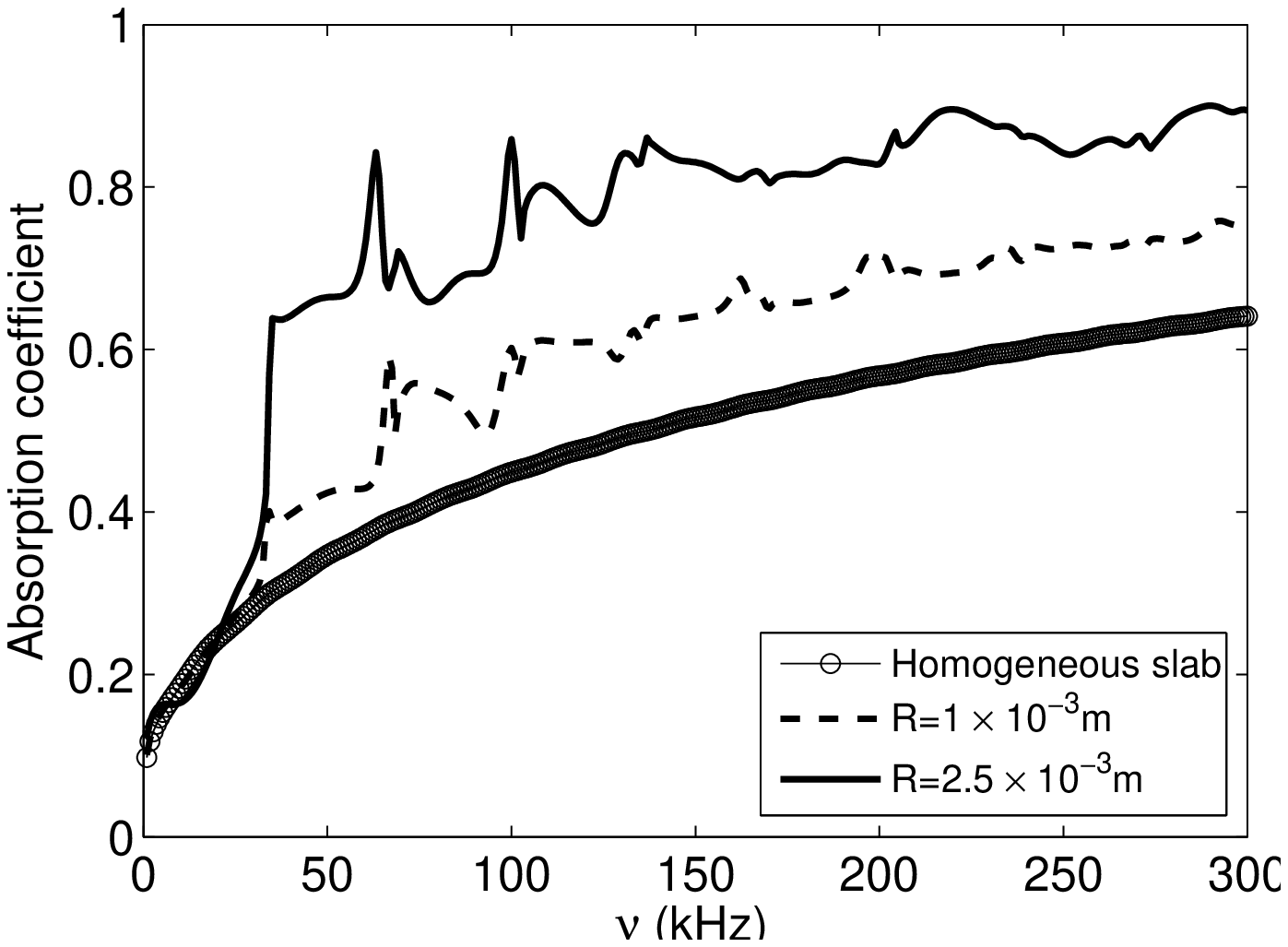,width=4.7cm}
\end{minipage}
\caption{Comparison of the reflection, transmission and absorption
coefficients for $R=2.5 \times 10^{-3}m$ and $R=1 \times 10^{-3}m$
type 3 inclusions when $d=10 \times 10^{-3}m$.} \label{section9f3}
\end{figure}

The developed form of the conservation of energy relation takes the form of
\begin{equation}
1=\mathcal{A}+\mathcal{R}+\mathcal{T}~,
\end{equation}
with $\mathcal{A}$, $\mathcal{R}$ and $\mathcal{T}$ the
absorption, hemispherical reflection and hemispherical
transmission coefficients  respectively, defined by
\begin{equation}
\begin{array}{l}
\displaystyle \mathcal{R}=\sum_{p\in \mathbb{Z}}\frac{\Re\left(k_{2p}^0\right)}{k_2^{0i}}\| R_p(\omega)\|^2=\sum_{p=-\bar{p}}^{\bar{p}}\frac{k_{2p}^0}{k_2^{0i}}\| R_p(\omega)\|^2~,\\
\displaystyle \mathcal{T}=\sum_{p\in\mathbb{Z}}\frac{\Re\left(k_{2p}^0\right)}{k_2^{0i}}\| T_p(\omega)\|^2=\sum_{p=-\bar{p}}^{\bar{p}}\frac{k_{2p}^0}{k_2^{0i}}\| T_p(\omega)\|^2~,
\end{array}
\end{equation}
wherein $\bar{p}$ is such that $\displaystyle \left(k_1^i+\frac{2\pi \left(\bar{p}+1 \right)}{d} \right)^2> (k^0)^2\geq \left(k_1^i+\frac{2\pi\bar{p}}{d} \right)^2 $.\\
\begin{figure}[H]
\begin{minipage}{7.0cm}
\centering\psfig{figure=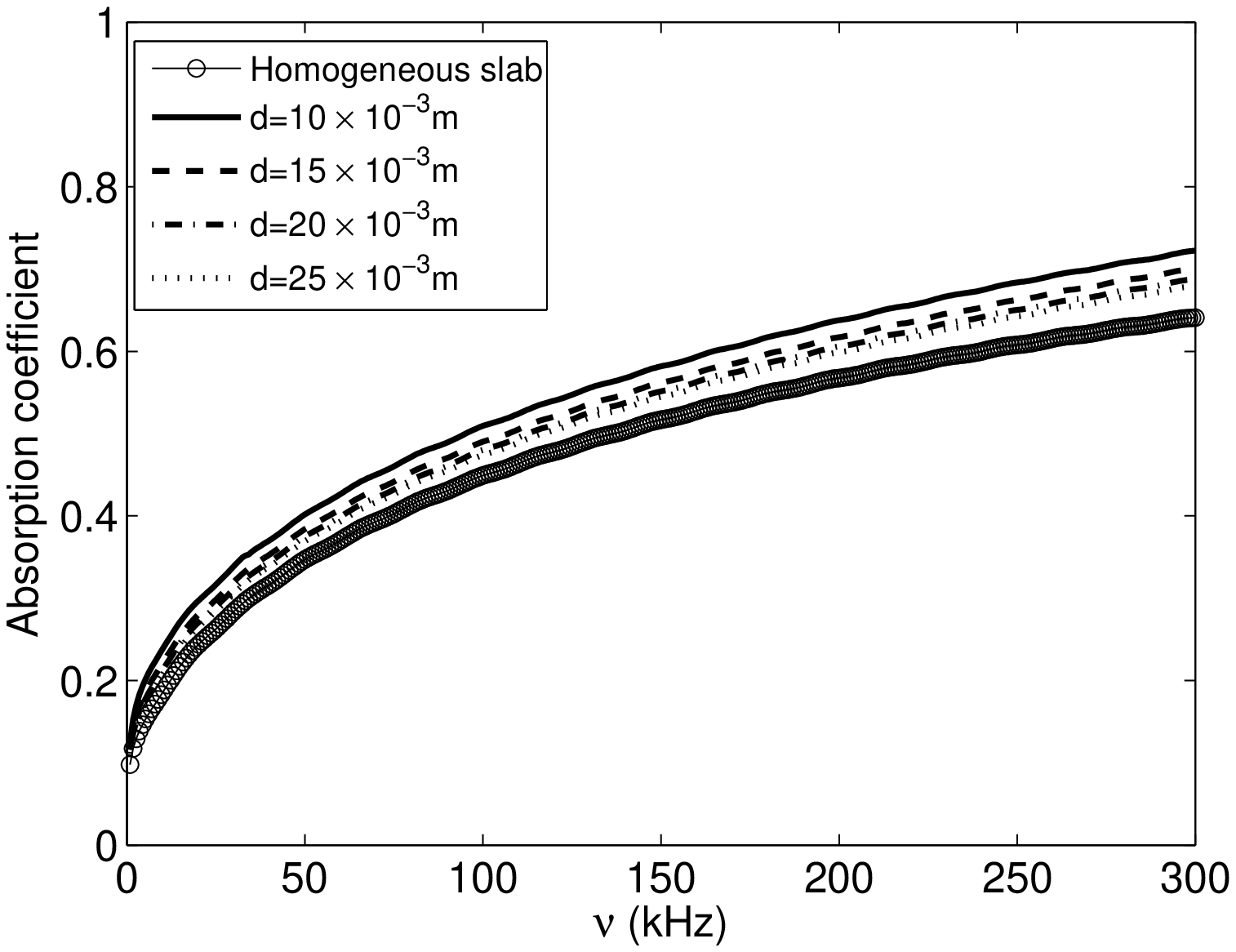,width=7.0cm}
\end{minipage}\hfill
\begin{minipage}{7.0cm}
\centering\psfig{figure=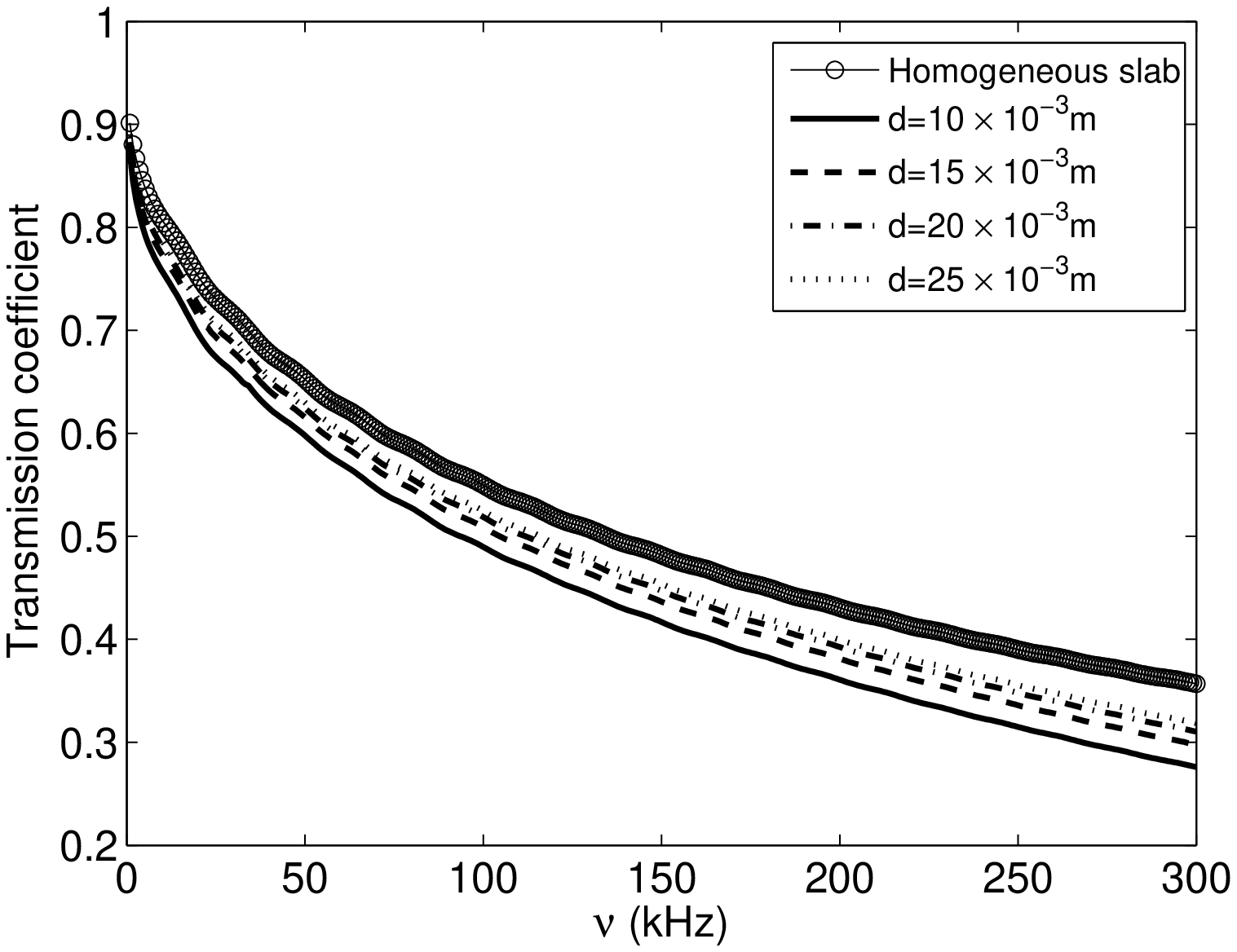,width=7.0cm}
\end{minipage}
\begin{minipage}{7.0cm}
\centering\psfig{figure=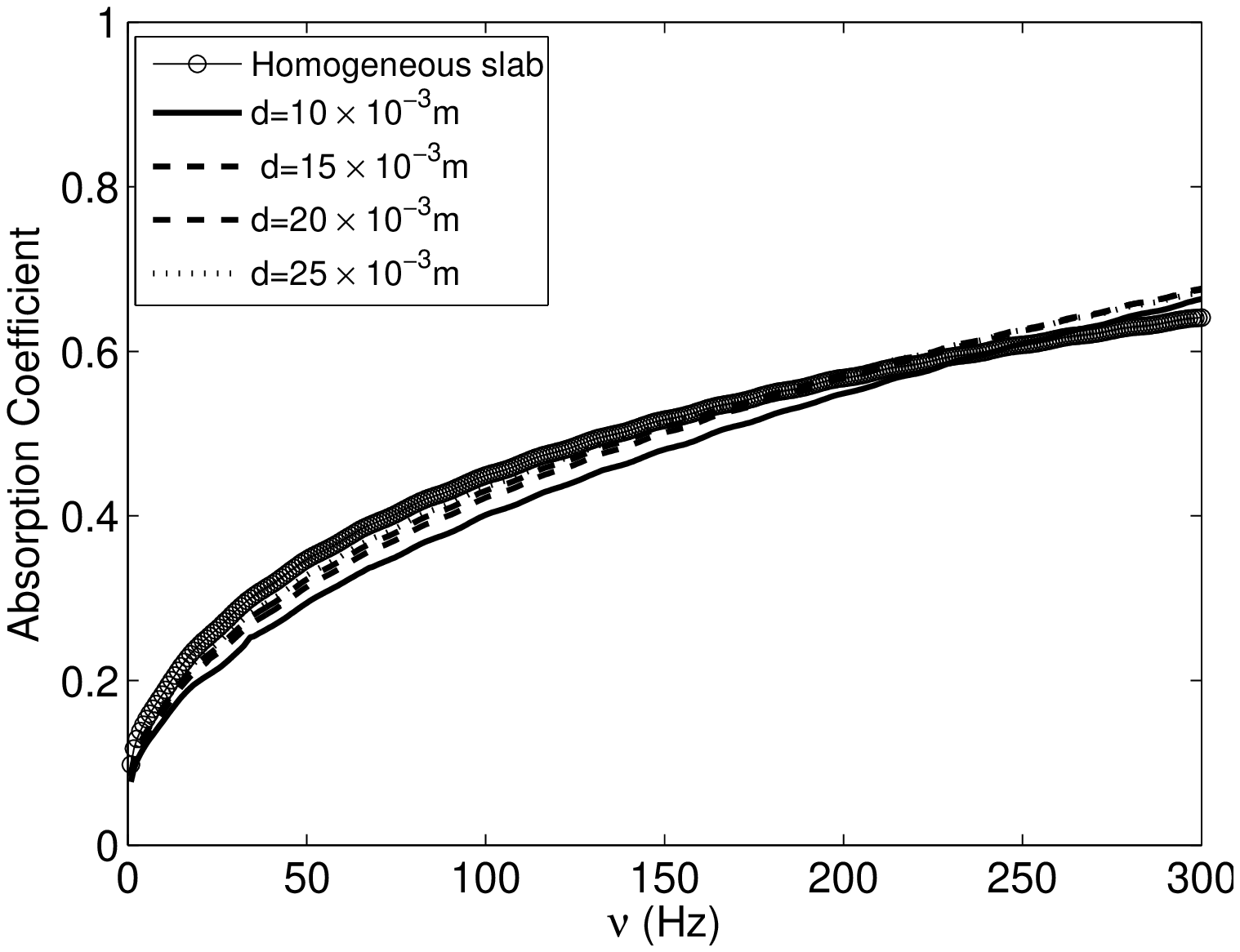,width=7.0cm}
\end{minipage}\hfill
\begin{minipage}{7.0cm}
\centering\psfig{figure=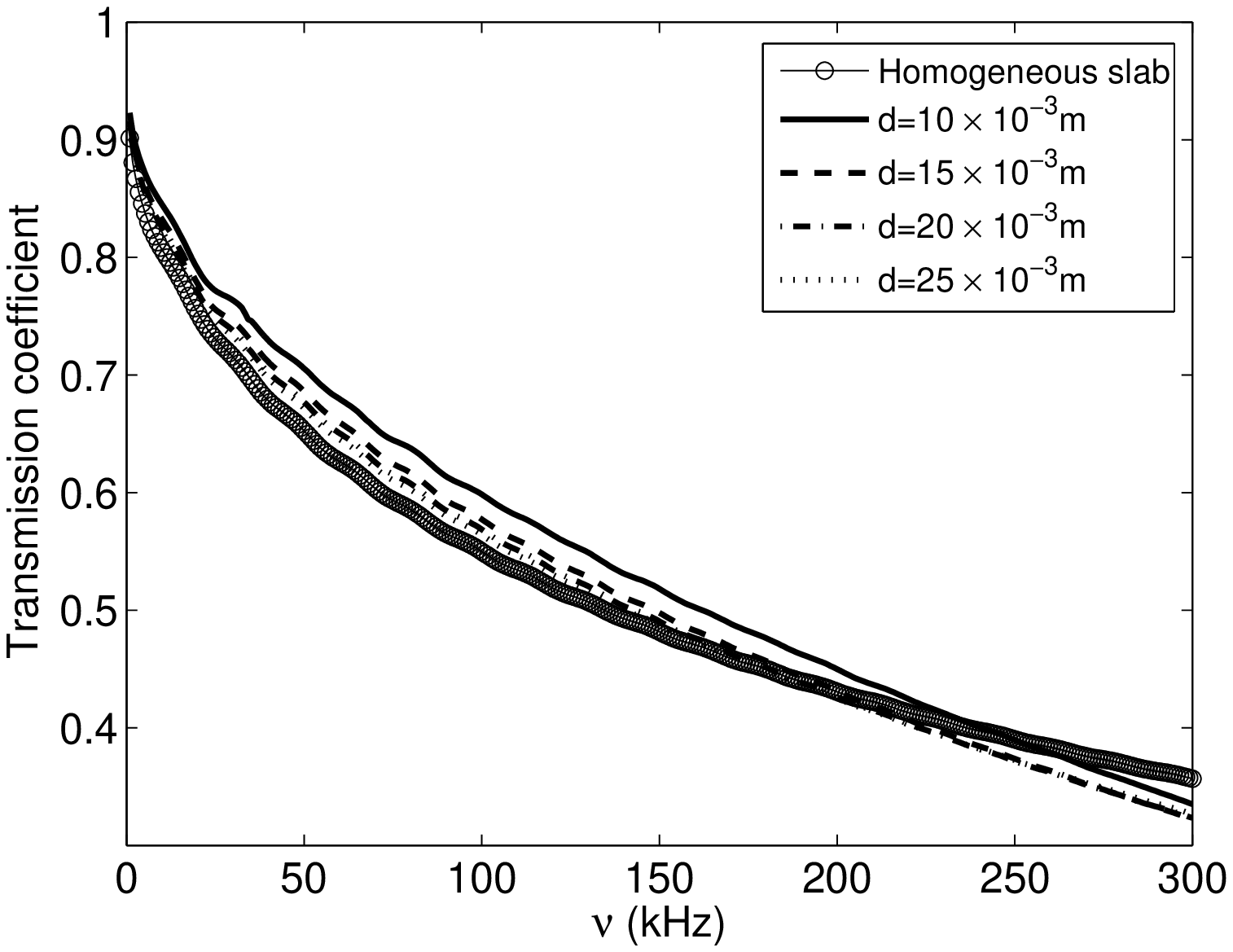,width=7.0cm}
\end{minipage}
\begin{minipage}{7.0cm}
\centering\psfig{figure=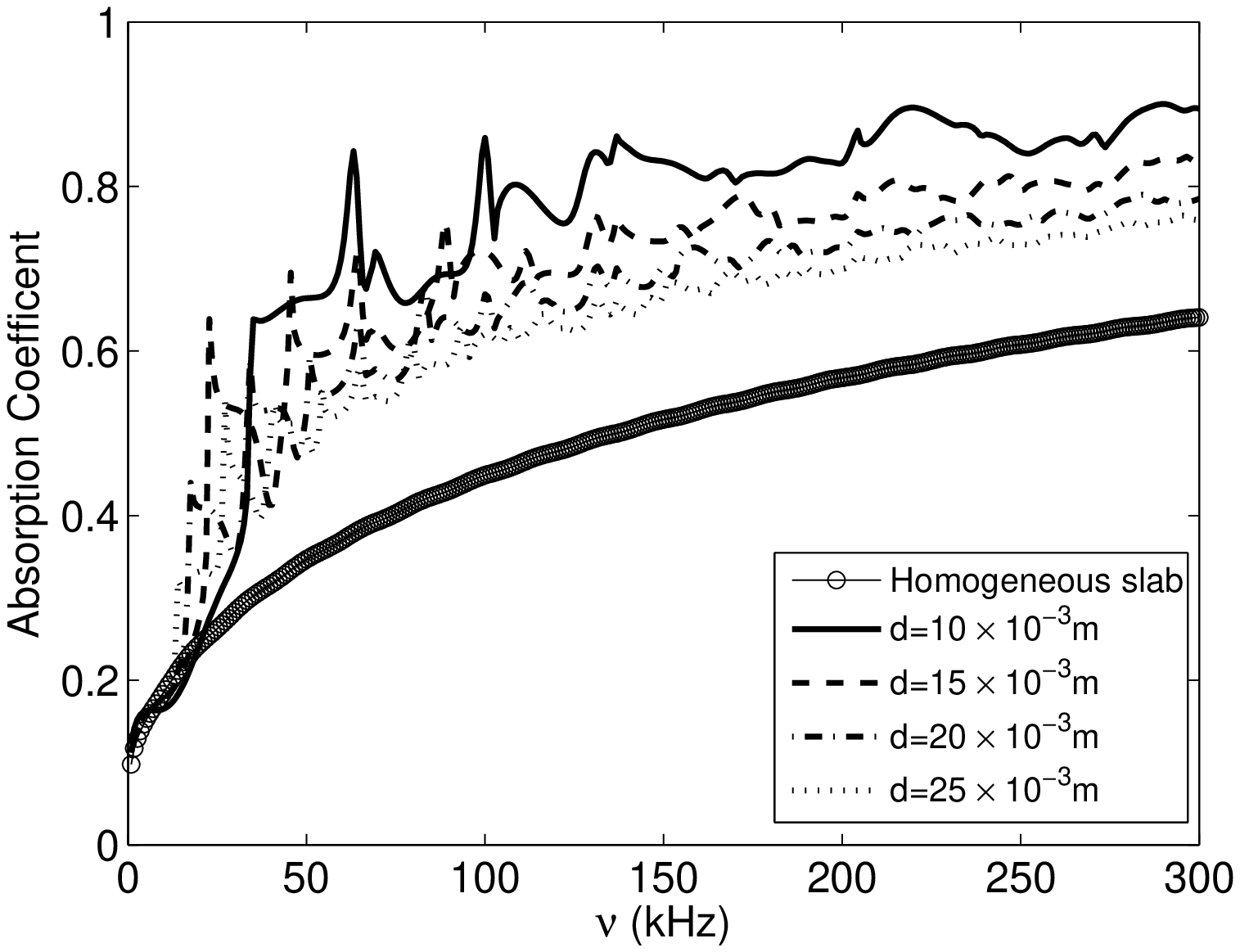,width=7.0cm}
\end{minipage}\hfill
\begin{minipage}{7.0cm}
\centering\psfig{figure=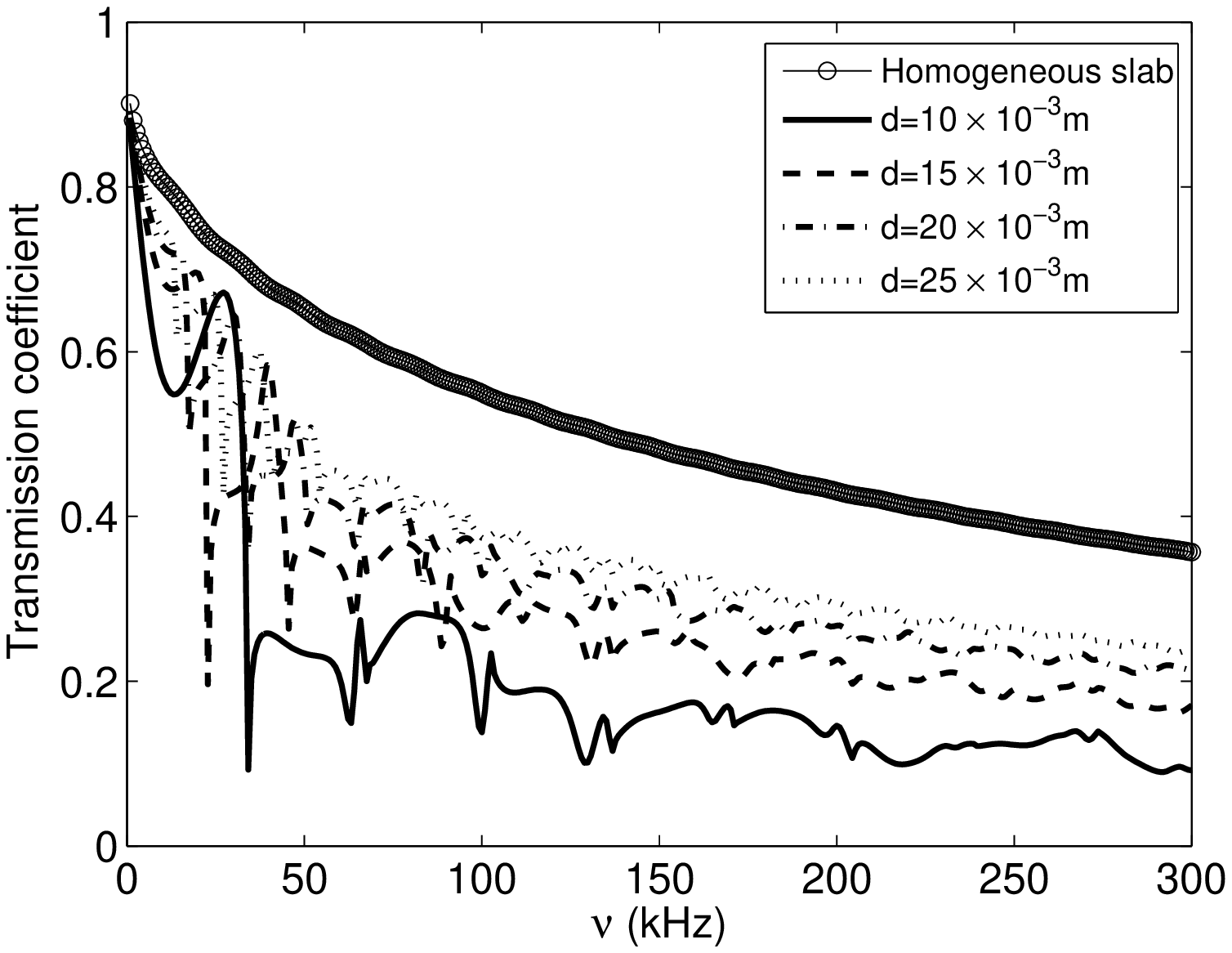,width=7.0cm}
\end{minipage}
\caption{Comparison of the absorption (left panel) and
hemispherical transmission (right panel) coefficients between the
cases of inclusions of type referenced by 1 (top panels), 2
(middle panels) or 3 (bottom panels) for various center-to center
distances.} \label{section9f1}
\end{figure}
The reflection coefficients are almost the same  when compared
with those as calculated in absence of inclusions except for
closed inclusions of type 3. The increase of the absorption
coefficient is also due to the decrease of the transmission
coefficient, as was previously noticed in \cite{tournat}. In all
the cases, the larger is the spatial period, the closer the
coefficients are to those of the case of a homogeneous slab.\\
\begin{figure}[H]
\centering\psfig{figure=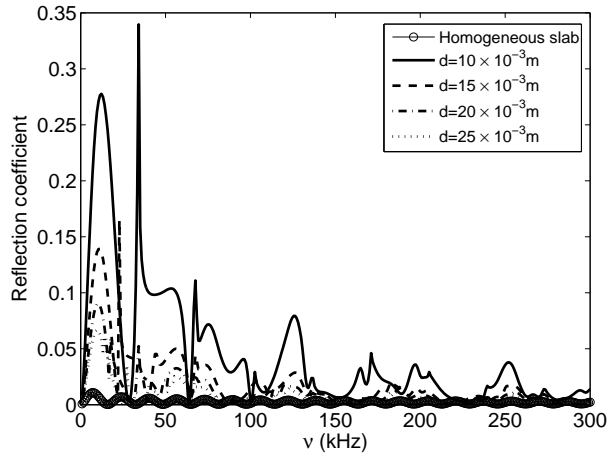,width=8.0cm}
\caption{Comparison of the reflection coefficient with inclusions
of type  3 for various center-to center distance.}
\label{section9f2}
\end{figure}
Fig.  \ref{section9f3} compares  the reflection, transmission and
absorption coefficients for $R=2.5 \times 10^{-3}m$ and $R=1
\times 10^{-3}m$ type 3 inclusions when $d=10\times 10^{-3}m$. The
smaller the inclusions are, the closer the coefficients are to
those of the case of the homogeneous slab.
\section{Discussion}
The addition of weak contrast inclusions (i.e. type 1 and  type 2)
leads to an absorption spectrum whose shape is close to that of a
macroscopically-homogeneous slab. We also deal with an effective
homogeneous porous slab, whose absorption coefficient is lower
when the inclusions are filled with the air medium and larger when
the inclusions are filled with an other, more absorptive, fluid-like
porous medium. For fluid-like porous inclusions, the slope of the
absorption coefficient is nearly the same as it is for the
macroscopically-homogeneous slab, while for air inclusions (holes)
the slope is larger. That is why at low frequency the absorption
coefficient is lower than the one as calculated for the
homogeneous slab and becomes larger for a higher frequency.

The addition of  high-contrast inclusions (i.e. type 3) leads to
an absorption spectrum that is, on the average, larger than for
the macroscopically-homogeneous slab, and which presents some
additional peaks. The slope of the absorption spectrum is closer
to that of the
 the macroscopically-homogeneous slab. Each peak leads to a
supplementary increase of the absorption of the configuration.  The
absorption can be multiplied by a factor of three at the location of
some of these peaks when compared with the one of the
macroscopically-homogeneous slab.  Two types of peaks are
distinguished. Each of them appears periodically for a specific
period of the inclusions and can be associated with mode excitation
of the whole configuration.

Two peaks stand out for all  center-to-center  distances at the same
frequencies near $\nu= 65 kHz$, $\nu= 133 kHz$. This corresponds to
the excitation of quasi-mode 2 and 3 of the initial slab . A
quasi-mode is a mode of the global configuration whose structure is
close to the one of a mode of a sub-structure it is composed of.
Quasi-mode 1 occurs for a frequency too low for it to be seen, and
quasi-mode 3 is largely attenuated. The excitation of the quasi-mode
of the slab is made possible by the spatial periodicity of the
configuration.

In the absence of the inclusions, the modes of a homogeneous slab
filled with a fluid-like porous medium cannot be excited by an
incident plane  wave  propagating in the air \cite{grobysapem}. If
$det(\mathbf{E})=0$ denotes a generic dispersion relation, then we
can say that $det(\mathbf{E})$ is close to zero (vanishes in absence
of dissipation) only for an evanescent wave in the air medium, which
cannot be excited by an incident plane bulk wave.

The spatially-periodic configuration leads to a field
representation, through the Floquet theorem, that includes
evanescent waves  the air medium. Another  explanation of the
quasi-mode excitation relies on the fact that each inclusion acts as
an induced cylindrical source. The response of an active cylindrical
source radiating in the neighborhood of a homogeneous slab (see
\cite{grobyMMAS2006}) enables a  mode of this slab to be excited
because some of the waves radiated by the source are evanescent and
have the same structure (at resonance) as that of the evanescent
wave associated with the mode.

The other peaks appear in a periodic manner beyond the peak
relative to the quasi-mode of the  slab. Their periodicity is
inversely proportional to the spatial periodicity $d$ of the
grating. This phenomena was already encountered in
\cite{grobyGJIII2006} and was attributed to a periodization of the
quasi-mode due to the primitive reciprocal lattice vector
$\displaystyle \mathbf{k_1}=\frac{2 \pi}{d} \mathbf{i_1}$. This
periodization is strongly associated with the excitation of quasi
Cutler-modes.

The increase  of the absorption is  also due to an average (global)
increase which does not take the form of additional peaks. This
global increase can be explained by multipathing between each
inclusion and/or by excitation of evanescent waves in the slab,
thanks to the existence of the grating.

Addition of high contrast inclusions also leads to an increase of
the absorption  of the slab, largely associated with a decrease of
the transmission coefficient. This is mainly due to mode excitation
of the whole configuration, whose structure consists of evanescent
waves in the air medium (and in the slab), thus leading to an
entrapment of the energy in the slab. The latter is dissipated by
thermal and viscous effects.

In all the cases, the  closer the inclusions are, and/or the larger
are their radii, the larger is the absorption.
\section{Conclusion}
We show that high-contrast, periodically-arranged, inclusions in a
porous slab induce  an increase of the absorption coefficient,
mainly associated with a decrease of the transmission coefficient
over a large frequency range, and in particular, at low frequencies
(although higher than $\sim 20kHz$). This effect is due to mode
excitation of the slab through excitation of the quasi-mode of the
initial slab (enabled) by the periodic inclusions, and to excitation
of grating modes via multipathing between the inclusions.

This increase of absorption is most noticeable for rigid frame
porous inclusions in a large portion of the  frequency range, and
is less pronounced for air inclusions.

The reflection coefficients  are found to be of the same order as
those in the absence of inclusions  for low-contrast inclusions,
and to be higher  than those in the absence of inclusions for high
contrast inclusions. The way of reducing the reflection is by
acting on the surface geometry of the slab. A first approach can
consist in the addition of a homogenized layer of double porosity
\cite{olny}.
\section*{Acknowledgment} The authors are garteful to D. Lesselier for his useful comments on earlier version of this paper. 
\appendix
\section{Numerical validation}
\begin{figure}[H]
\begin{minipage}{7.0cm}
\centering\psfig{figure=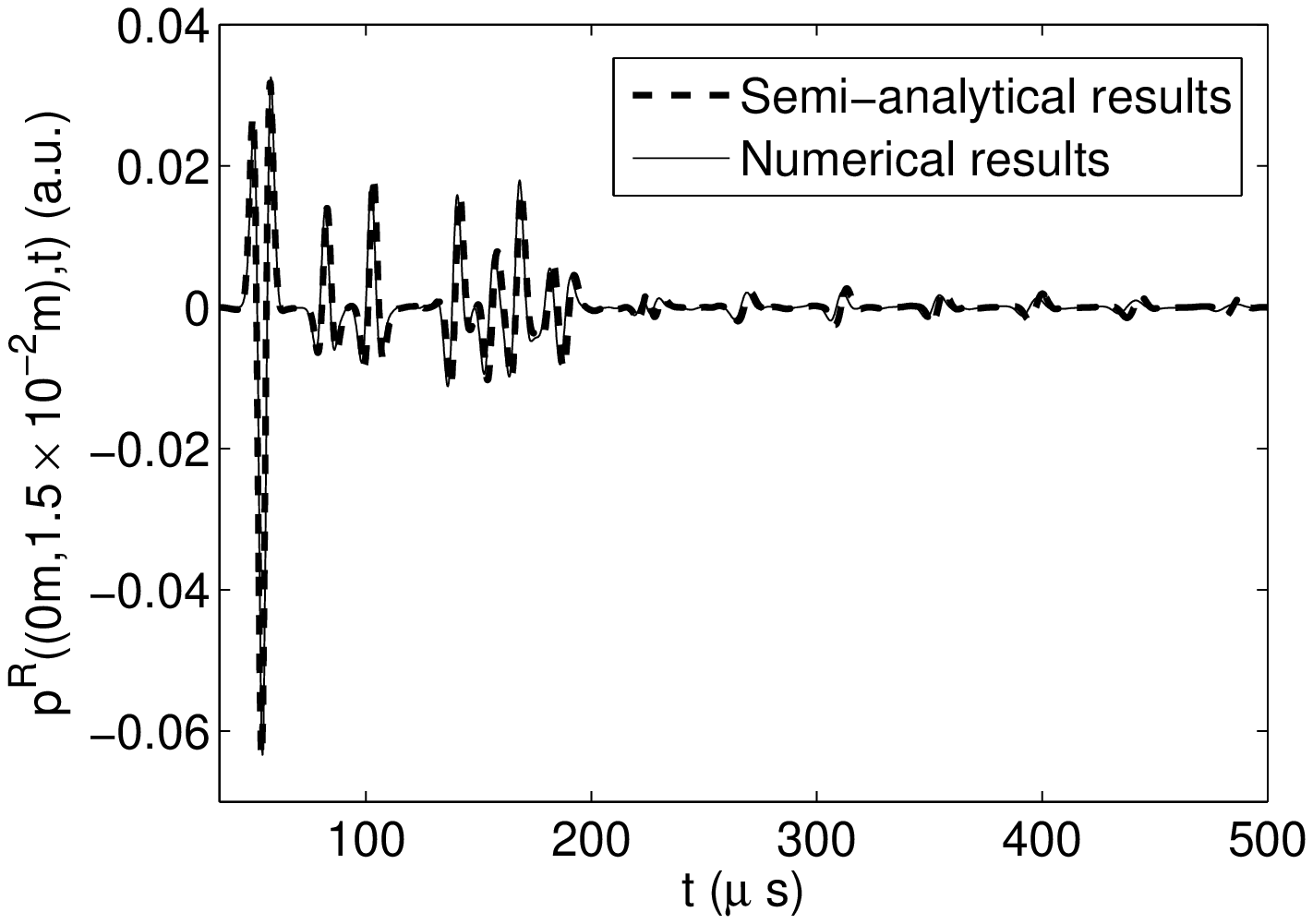,width=7.0cm}
\end{minipage}\hfill
\begin{minipage}{7.0cm}
\centering\psfig{figure=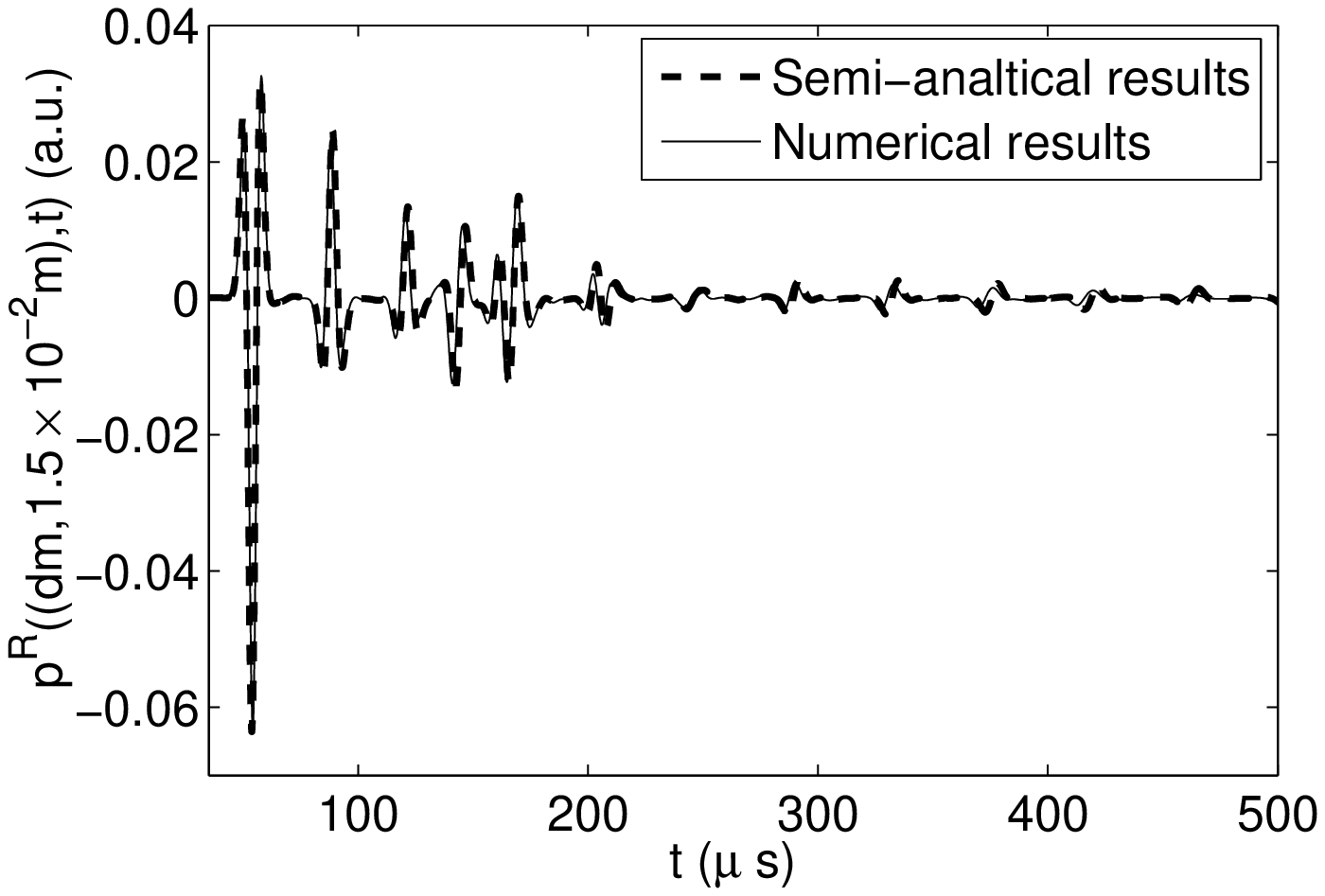,width=7.0cm}
\end{minipage}
\begin{minipage}{7.0cm}
\centering\psfig{figure=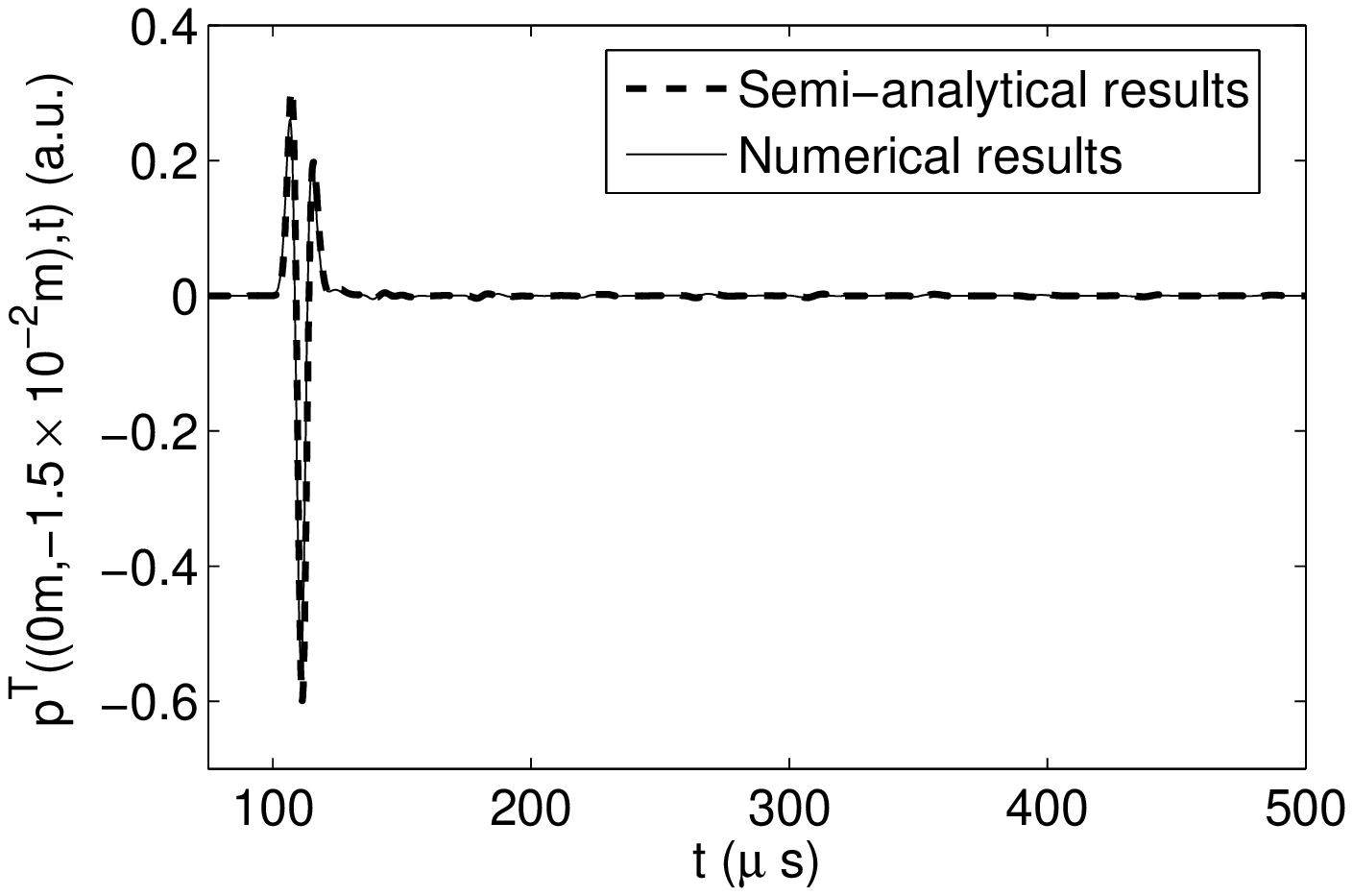,width=7.0cm}
\end{minipage}\hfill
\begin{minipage}{7.0cm}
\centering\psfig{figure=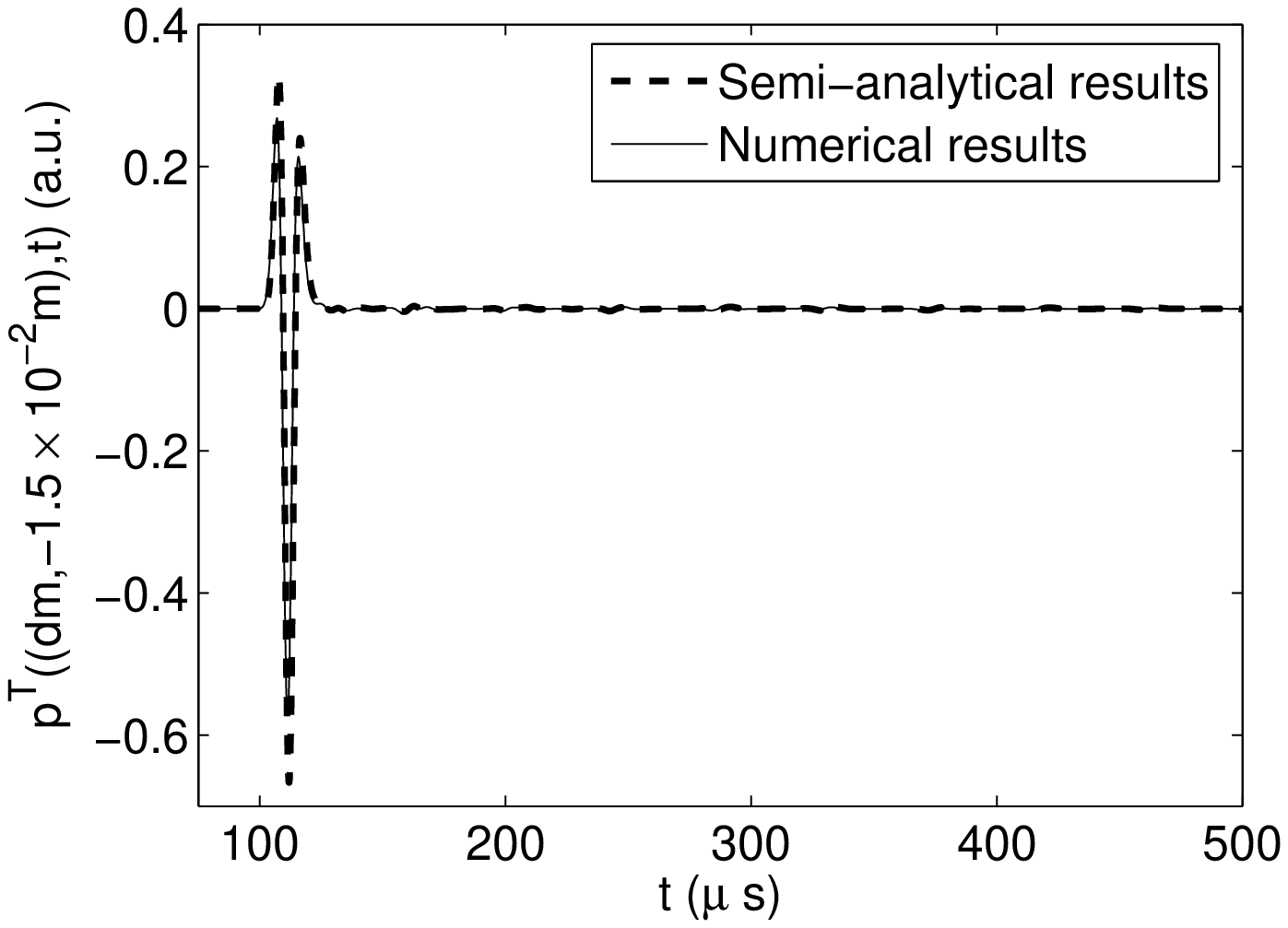,width=7.0cm}
\end{minipage}
\caption{Comparison between the reflected and transmitted pressure field as
calculated by the semi-analytical method and as calculated by the FE code. The
top panels depict the reflected field, while the bottom panels depict the transmitted
field for both a central location (left panels) and appendicular location (right panels) of the unit cells.}
\label{apfig1}
\end{figure}

We validate the numerical implementation of the analytical
calculation with the help of our Finite-Element code
\cite{grobyJCA}, in the case of a viscoacoustic medium (Fig. \ref{apfig1}).

The medium filling $\Omega_0$ is the air medium ($\rho_0=\rho_f$,
$\displaystyle c_0=\sqrt{\frac{\gamma P_0}{\rho_0}}$). The infinite slab and the inclusion are filled with
viscoacoustic media whose relaxed characteristics are those of the
porous medium when the dissipative aspects vanish, (i.e.
$\displaystyle \rho_1=\frac{c_0 \times 0.96}{1.07}$, $\displaystyle c_1=\frac{c_0}{\sqrt{1.07}}$
and $\displaystyle \rho_2=\frac{\rho_0 \times 0.99}{1.001}$,
$\displaystyle c_2=\frac{c_0}{\sqrt{1.001}}$). The dissipative aspects of
 $M^1$ and $M^2$ are described by constant quality
factors over the frequency range of solicitation, such that $Q^1=30$
and $Q^1=50$.

The numerical simulations are  performed on a grid of $3000 \times
400$ nodes, whose grid spacing is equal to $1 \times 10^{-4}m$. The
slab is $2 \times 10^{-2}m$ thick and the center-to-center distance
between two adjacent cylinders is $1.5 \times 10^{-2}m$. The radius
of each cylinder is $R=5 \times 10^{-3}m$.

The signal spectrum of the solicitation is that of a  Ricker pulse
centered at $\nu_0=100 kHz$. The incident angle is $\theta^{i}=0$.
\bibliography{biblio_an}
\bibliographystyle{plain}
\end{document}